\documentclass[11pt]{article}
\pdfoutput=1
\usepackage{jheppub}
\usepackage{graphicx}
\usepackage{amssymb}
\usepackage{amsmath,amssymb}
\usepackage{slashed}
\usepackage{hyperref}
\usepackage{caption}
\usepackage{xcolor}
\usepackage{dsfont}
\usepackage{verbatim}
\usepackage{subfig}
\usepackage{amsmath,amssymb,amscd,amsfonts,mathtools,csquotes}
\usepackage{xifthen}
\usepackage{xcolor}
\definecolor{markgreen}{RGB}{230,243,230}
\definecolor{darkolivegreen}{rgb}{0.33, 0.42, 0.18}
\definecolor{darkpastelgreen}{rgb}{0.01, 0.75, 0.24}

\usepackage[toc,page]{appendix}
\usepackage{epsfig}
\usepackage{epstopdf}
\usepackage{latexsym}
\usepackage{graphicx}
\usepackage{booktabs}
\usepackage{bbm}
\usepackage{color}
\usepackage{physics}
\usepackage{tensor}
\usepackage{verbatim}
\usepackage{subfig}
\usepackage{tikz}
\usepackage{ifthen}
\usetikzlibrary{matrix}
\usetikzlibrary{decorations.markings,calc,shapes,decorations.pathmorphing,patterns,decorations.pathreplacing,arrows.meta}
\usetikzlibrary{positioning}
\usepackage{hyperref}
\usetikzlibrary{patterns}
\usetikzlibrary{positioning}

\pdfoutput=1
\makeatletter
\def\@fpheader{\relax}
\newcommand*{\ov}[1]{%
  $\m@th\overline{\mbox{#1}}$%
}
\newcommand*{\ovA}[1]{%
  $\m@th\overline{\mbox{#1}\raisebox{3mm}{}}$%
}
\newcommand*{\ovB}[1]{%
  $\m@th\overline{\mbox{#1\rule{0pt}{3mm}}}$%
}
\newcommand*{\ovC}[1]{%
  $\m@th\overline{\mbox{#1\strut}}$%
}
\newcommand*{\ovD}[1]{%
  $\m@th\overline{\mbox{#1\vphantom{\"A}}}$%
}
\newcommand*{\ovE}[1]{%
  $\m@th\overline{\raisebox{0pt}[1.2\height]{#1}}$%
}
\newcommand*{\ovF}[1]{%
  $\m@th\overline{\raisebox{0pt}[\dimexpr\height+1mm\relax]{#1}}$%
}
\newcommand*{\ovG}[1]{%
  $\m@th\overline{\raisebox{0pt}[\dimexpr\height+1mm\relax]{#1\vphantom{A}}}$%
}
\makeatother
\newboolean{showcomments}
\setboolean{showcomments}{false}
\newcommand\rem[1]{\ifthenelse{\boolean{showcomments}}{{#1}}{}}
\newcommand{\be}{\begin{equation}}
\newcommand{\ee}{\end{equation}}

\usepackage{xifthen}
\newcommand{\dalembert}[1][]{\ifthenelse{\isempty{#1}}{\Box}{#1\Box}}
\usetikzlibrary{decorations.pathmorphing}
\tikzset{snake it/.style={decorate, decoration=snake}}

\title{\Large The Mechanism behind the Information Encoding for Islands}

\author{Hao Geng}
\affiliation{Gravity, Spacetime, and Particle Physics (GRASP) Initiative, Harvard University, 17 Oxford St., Cambridge, MA, 02138, USA.}

\emailAdd{haogeng@fas.harvard.edu}

\abstract{Entanglement islands are subregions in a gravitational universe whose information is fully encoded in a disconnected non-gravitational system away from it. In the context of the black hole information paradox, entanglement islands state that the information about the black hole interior is encoded in the early-time Hawking radiation. Nevertheless, it was unclear how this seemingly nonlocal information encoding scheme emerges from a manifestly local theory. In this paper, we provide an answer to this question by uncovering the mechanism behind this information encoding scheme. As we will see, the early understanding that graviton is massive in island models plays an essential role in this mechanism. As an example, we will discuss how this mechanism works in detail in the Karch-Randall braneworld. This study also suggests the potential importance of this mechanism to the ER=EPR conjecture.}

\begin{document}
\maketitle
\flushbottom
\pagebreak

\section{Introduction}
Entanglement islands emerged in the attempt to compute the Page curve of the radiation from evaporating black holes in anti-de Sitter (AdS) spacetime coupled to an auxiliary non-gravitational bath \cite{Almheiri:2019psf,Penington:2019npb}. Interestingly it was noticed later that entanglement islands are in fact ubiquitous in the setup with a gravitational asymptotically AdS spacetime coupled to an auxiliary non-gravitational bath, which we will call \textit{the island model}, even without a black hole \cite{Almheiri:2019yqk}. This suggests that entanglement islands are elementary in quantum gravity, not specifically to evaporating black holes, and hence deserve a better understanding.

\begin{figure}
    \centering
    \subfloat[Island Model with a Black Hole \label{pic:penroseoriginal}]
{
    \begin{tikzpicture}[scale=0.65,decoration=snake]
       \draw[-,very thick] 
       decorate[decoration={zigzag,pre=lineto,pre length=5pt,post=lineto,post length=5pt}] {(-2.5,0) to (2.5,0)};
       \draw[-,very thick,red] (-2.5,0) to (-2.5,-5);
       \draw[-,very thick,red] (2.5,0) to (2.5,-5);
         \draw[-,very thick] 
       decorate[decoration={zigzag,pre=lineto,pre length=5pt,post=lineto,post length=5pt}] {(-2.5,-5) to (2.5,-5)};
       \draw[-,very thick] (-2.5,0) to (2.5,-5);
       \draw[-,very thick] (2.5,0) to (-2.5,-5);
       \draw[-,very thick,green] (-2.5,0) to (-5,-2.5);
       \draw[-,very thick,green] (-5,-2.5) to (-2.5,-5);
        \draw[-,very thick,green] (2.5,0) to (5,-2.5);
       \draw[-,very thick,green] (5,-2.5) to (2.5,-5);
       \draw[fill=green, draw=none, fill opacity = 0.1] (-2.5,0) to (-5,-2.5) to (-2.5,-5) to (-2.5,0);
       \draw[fill=green, draw=none, fill opacity = 0.1] (2.5,0) to (5,-2.5) to (2.5,-5) to (2.5,0);
       \draw[->,very thick,black] (-2.2,-3.5) to (-2.2,-1.5);
       \node at (-1.8,-2.5)
       {\textcolor{black}{$t$}};
        \draw[->,very thick,black] (2.2,-3.5) to (2.2,-1.5);
       \node at (1.8,-2.5)
       {\textcolor{black}{$t$}};
       \draw[-,thick, blue] (-5,-2.5) to (5,-2.5);
       \draw[-,thick,red] (-5,-2.5) to (-3.5,-2.5);
       \draw[-,thick,red] (5,-2.5) to (3.5,-2.5);
       \draw[-,thick,blue] (5,-2.5) arc (60:75.5:10);
       \draw[-,thick,blue] (-5,-2.5) arc (120:104.5:10);
       \draw[-,thick,blue] (2.5,-1.47) to (0,-2.5);
       \draw[-,thick,blue] (-2.5,-1.47) to (0,-2.5);
       \draw[-,thick,red] (5,-2.5) arc (60:70:10);
       \draw[-,thick,red] (-5,-2.5) arc (120:110:10);
       \node at (-3.8,-2.2)
       {\textcolor{red}{$R_{I}$}};
        \node at (3.8,-2.2)
       {\textcolor{red}{$R_{II}$}};
       \draw[->,decorate,orange] (-2.9,-4.4) to (-2,-3.4);
        \draw[->,decorate,orange] (-2,-1.5) to (-2.9,-0.6);
         \draw[->,decorate,orange] (2.9,-4.4) to (2,-3.4);
        \draw[->,decorate,orange] (2,-1.5) to (2.9,-0.6);
    \end{tikzpicture}}
    \hspace{1.0cm}
    \subfloat[Island Rule \label{pic:penroseisland}]
{
    \begin{tikzpicture}[scale=0.65,decoration=snake]
       \draw[-,very thick] 
       decorate[decoration={zigzag,pre=lineto,pre length=5pt,post=lineto,post length=5pt}] {(-2.5,0) to (2.5,0)};
       \draw[-,very thick,red] (-2.5,0) to (-2.5,-5);
       \draw[-,very thick,red] (2.5,0) to (2.5,-5);
         \draw[-,very thick] 
       decorate[decoration={zigzag,pre=lineto,pre length=5pt,post=lineto,post length=5pt}] {(-2.5,-5) to (2.5,-5)};
       \draw[-,very thick] (-2.5,0) to (2.5,-5);
       \draw[-,very thick] (2.5,0) to (-2.5,-5);
       \draw[-,very thick,green] (-2.5,0) to (-5,-2.5);
       \draw[-,very thick,green] (-5,-2.5) to (-2.5,-5);
        \draw[-,very thick,green] (2.5,0) to (5,-2.5);
       \draw[-,very thick,green] (5,-2.5) to (2.5,-5);
       \draw[fill=green, draw=none, fill opacity = 0.1] (-2.5,0) to (-5,-2.5) to (-2.5,-5) to (-2.5,0);
       \draw[fill=green, draw=none, fill opacity = 0.1] (2.5,0) to (5,-2.5) to (2.5,-5) to (2.5,0);
       \draw[-,thick,blue] (5,-2.5) arc (60:75.5:10);
       \draw[-,thick,blue] (-5,-2.5) arc (120:104.5:10);
       \draw[-,thick,blue] (2.5,-1.47) to (1.25,-1.985);
       \draw[-,very thick,purple!!!] (1.25,-1.985) to (0,-2.5);
       \draw[-,thick,blue] (-2.5,-1.47) to (-1.25,-1.985);
       \draw[-,very thick,purple!!!] (-1.25,-1.985) to (0,-2.5);
         \node at (0,-2.)
       {\textcolor{purple}{$\mathcal{I}$}};
       \draw[-,thick,red] (5,-2.5) arc (60:70:10);
       \draw[-,thick,red] (-5,-2.5) arc (120:110:10);
       \node at (-3.8,-2.2)
       {\textcolor{red}{$R_{I}$}};
        \node at (3.8,-2.2)
       {\textcolor{red}{$R_{II}$}};
       \draw[->,decorate,orange] (-2.9,-4.4) to (-2,-3.4);
        \draw[->,decorate,orange] (-2,-1.5) to (-2.9,-0.6);
         \draw[->,decorate,orange] (2.9,-4.4) to (2,-3.4);
        \draw[->,decorate,orange] (2,-1.5) to (2.9,-0.6);
    \end{tikzpicture}}
   
    \caption{\small \textbf{a)} The Penrose diagram of the island model with a black in the AdS$_{d+1}$. The two red vertical lines denotes the conformal boundary of the AdS$_{d+1}$ black hole. The green shaded regions are the nongravitational bath whose geometry is the flat Minkowski space. The orange arrows denotes the radiation coming in and out of the black hole. Under the time evolution chosen in the diagram, more and more radiation from the black hole will be captured by the bath. Two Cauchy slices of this time evolution are denoted by the two blue curves and on each of them we specify the subsystem $R=R_{I}\cup R_{II}$ as the red intervals. \textbf{b)} A putative configuration with entanglement island. The island is denoted as the purple interval in the black hole spacetime. The causal diamond of the island overlaps the black hole interior which is the reason the interior of the black hole is encoded in the bath.}
\end{figure}
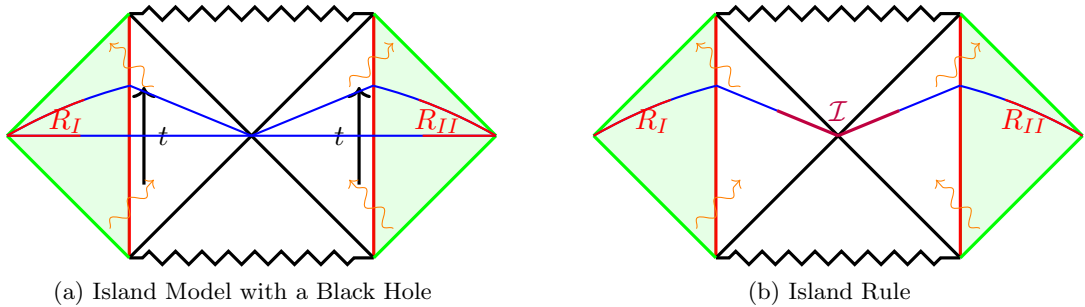

In the island model, one is looking for the entanglement wedge of a bath subregion $R$. Interestingly, its entanglement wedge can contain a subregion $\mathcal{I}$ in the gravitational AdS which is disconnected from $R$. This AdS subregion $\mathcal{I}$ is called the entanglement island of the bath subregion $R$. The island can be located using the following \textit{island formula}
\begin{equation}
    S(R)=\min_{ \mathcal{I}}\Big[S_{\text{QFT}}(R\cup\mathcal{I})+\frac{A(\partial\mathcal{I})}{4G_{N}}\Big]\,,\label{eq:islandformula}
\end{equation}
where $S(R)$ is the entanglement entropy of the subregion $R$, $S_{\text{QFT}}(R\cup \mathcal{I})$ is the entanglement entropy for the quantum fields in the subregion $R\cup \mathcal{I}$ in a fixed non-gravitational background, $A(\partial\mathcal{I})$ is the area of the boundary of the island $\mathcal{I}$ and $G_{N}$ is the AdS Newton's constant. As an output of the formula Equ.~(\ref{eq:islandformula}), one will know the value of the entanglement entropy of the bath subregion $R$ and the location of its entanglement island $\mathcal{I}$ inside the gravitational AdS. In the context of black holes, one can take $R$ as its early-time Hawking radiation collected by the bath (see Fig.\ref{pic:penroseoriginal}). Therefore, $S(R)$ is, in fact, the entanglement entropy of early-time Hawking radiation, and one can use Equ.~(\ref{eq:islandformula}) to compute the time dependence of this entropy \cite{Almheiri:2019psf,Penington:2019npb} which will give the unitary Page curve \cite{Page:1993wv}. In this calculation, the emergence of the unitary Page curve relies on the existence of a non-vanishing island at late time (see Fig.\ref{pic:penroseisland}). The statement that $\mathcal{I}$ is part of the entanglement wedge of $R$ indicates that the information inside the island is fully encoded in the bath subregion $R$.

Nevertheless, due to the fact that the early works \cite{Almheiri:2019psf,Penington:2019npb,Almheiri:2019yqk,Almheiri:2019qdq,Penington:2019kki} were done in low-dimensional models of quantum gravity, where dynamics of the gravitational field is almost trivial, it is only very recently being the physics behind the entanglement islands gradually understood \cite{Geng:2020qvw,Geng:2021hlu,Geng:2023ynk,Geng:2023zhq,Geng:2024xpj}. These progress are largely motivated by the physics of the Karch-Randall braneworld \cite{Karch:2000ct,Karch:2000gx,Geng:2023qwm}. The Karch-Randall braneworld provides a natural holographic dual of the island model. Due to its holographic nature, one can easily search for entanglement islands \cite{Almheiri:2019hni} even in higher spacetime dimensions \cite{Almheiri:2019psy,Geng:2020qvw}. This circumvents the difficulty that one doesn't have a formula for entanglement entropy of higher dimensional quantum field theory in a curved background to operate the island formula Equ.~(\ref{eq:islandformula}). In \cite{Geng:2020qvw}, motivated by the early understanding in \cite{Karch:2000ct,Karch:2000gx} that the graviton is massive in the higher dimensional Karch-Randall braneworld,\footnote{In fact it was understood that the AdS graviton is massive in higher dimensional island models \cite{Porrati:2001gx,Porrati:2003sa,Duff:2004wh,Aharony:2006hz,Geng:2023ynk}. Hence the massive graviton is not a specific feature of the Karch-Randall braneworld.} we conjectured that entanglement islands can exist only in massive gravity theories. Further evidence of this conjecture was found in \cite{Geng:2020fxl} that in a deformed version of the original Karch-Randall braneworld, with a massless graviton, there is in fact no entanglement island. A general proof of this conjecture for a large class of situations of physical interest was provided in \cite{Geng:2021hlu} by noticing that entanglement island, as an entanglement wedge which doesn't contain any portion of the AdS asymptotic boundary, is not consistent with the long-range nature of the standard massless gravity theory. This inconsistency is due to the fact that the entanglement island $\mathcal{I}$ is part of the entanglement wedge of the bath subregion $R$ and the complement of the entanglement island $\bar{\mathcal{I}}$ in the AdS, including its asymptotic boundary, is in the entanglement wedge of $\bar{R}$, the complement of $R$ in the bath. Hence operators in $\mathcal{I}$ should commute with operators in $\bar{\mathcal{I}}$ as operators in $R$ and $\bar{R}$. However, this is not consistent with the Gauss' law of the standard massless gravity that the Hamiltonian of the gravitational AdS is a boundary term accessible around the asymptotic boundary of the AdS. This boundary Hamiltonian is called the ADM Hamiltonian $H_{\text{ADM}}$ in the general relativity literature \cite{Arnowitt:1962hi}. Therefore,  in  the standard massless gravity theories any local operator in the AdS bulk wouldn't commute with $H_{\text{ADM}}$, including the operators in $\mathcal{I}$, and this is in contradiction with the above statement that operators in the island should commute with operators in its complement. The long-range property is broken in the massive gravity theories, due to the Yukawa type exponential decay of the gravitational potential, which therefore is the natural habitat for entanglement islands. Interestingly, the recent progress in \cite{Geng:2023zhq} has successfully generalized the above lessons to low-dimensional models used in the early works \cite{Almheiri:2019psf,Penington:2019npb,Almheiri:2019yqk,Almheiri:2019qdq,Penington:2019kki} by noticing that in higher dimensional island models the graviton mass is an indicator of the spontaneous breaking of the AdS diffeomorphism symmetry.  Moreover, studies in \cite{Geng:2023zhq} show that in the manifestly covariant description of this Higgs phase one can construct local operators in AdS that commute with the ADM Hamiltonian in both low-dimensional and high-dimensional island models. Hence a satisfying understanding of the question why information in the island is encoded in the non-gravitational bath but not the asymptotic boundary of the AdS as in the standard holographic quantum gravitational theory is obtained. However, it is still unclear how such an encoding scheme is achieved. This is puzzling as the information in the island seems to be encoded in the bath in a nonlocal way as they are spacelike separated and disconnected, so it is an important question to understand how such a seemingly nonlocal information encoding is achieved in a manifestly local theory.\footnote{See \cite{Giddings:2021qas,Martinec:2022lsb,Guo:2021blh} for other requests for a better understanding of this question.}

In this paper, we will address the above question by uncovering the mechanism for how the operators in the gravitational AdS in the island model are nontrivially evolved by the bath Hamiltonian. Even though this mechanism cannot answer fine-grained questions like what exactly the pattern of information encoding of islands by the bath is, its universality is a cornerstone and suggests potential relevance of it to fundamental questions like the ER=EPR conjecture \cite{Maldacena:2013xja}. The answer for those fine-grained questions like the one above is largely model-dependent but a general calculational algorithm that is able to answer them case by case in an exact way might exist.  We defer the search for such an algorithm to future work. Such an algorithm shall be the final quantum gravity theory which is likely the fully fledged string theory.

This paper is organized as follows. In Sec.~\ref{sec:review} we review the physics in the island model and formulate the question we are addressing in this paper in a precise way. In Sec.~\ref{sec:mechanism} we first provide more detailed studies of holography in the island which then enable us to uncover the mechanism behind the information encoding of island by the bath in the island model. In Sec.~\ref{sec:toymodel} we exam this mechanism in the Karch-Randall braneworld in detail where we will see that this mechanism nicely geometrizes and this suggests the potential importance of this mechanism to the ER=EPR conjecture. We conclude our paper with discussions in Sec.~\ref{sec:conclusions}. Potential questions and their answers are discussed in the Appendix.

\section{Physics in the Island Model}\label{sec:review}
In this section, we review essential physics in the island model as the background for our later discussions. For the sake of simplicity, we focus on the island model with no black holes. Most of the discussions in this section are based on the results of the earlier works \cite{Porrati:2001gx,Porrati:2003sa,Duff:2004wh,Aharony:2006hz,Geng:2021hlu,Geng:2023ynk,Geng:2023zhq}.

\subsection{An Explicit Island Model}\label{sec:islandmodel}
\begin{figure}
    \centering
    \begin{tikzpicture}
       \draw[-,very thick,red](0,-2) to (0,2);
       \draw[fill=green, draw=none, fill opacity = 0.1] (0,-2) to (4,-2) to (4,2) to (0,2);
           \draw[-,very thick,red](0,-2) to (0,2);
       \draw[fill=blue, draw=none, fill opacity = 0.1] (0,-2) to (-4,-2) to (-4,2) to (0,2);
       \node at (-2,0)
       {\textcolor{black}{$AdS_{d+1}$}};
        \node at (2,0)
       {\textcolor{black}{$AdS_{d+1}$}};
       \draw [-{Computer Modern Rightarrow[scale=1.25]},thick,decorate,decoration=snake] (-1,-1) -- (1,-1);
       \draw [-{Computer Modern Rightarrow[scale=1.25]},thick,decorate,decoration=snake] (1,1) -- (-1,1);
    \end{tikzpicture}
    \caption{We couple the gravitational AdS$_{d+1}$ universe (the blue shaded region) with a nongravitational bath (the green shaded region) by gluing them along the asymptotic boundary the of AdS$_{d+1}$ (the red vertical line). The nongravitational bath is modeled by another AdS$_{d+1}$ which shares the same asymptotic boundary with the original AdS$_{d+1}$. We take the Poincar\'{e} coordinates in both of the AdS$_{d+1}$. The coupling is achieved as described by Equ.~(\ref{eq:asympcouple}).}
    \label{pic:islandmodel}
\end{figure}
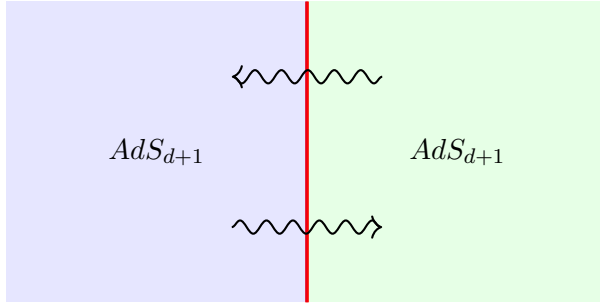

In the island model, a gravitational AdS$_{d+1}$ spacetime is coupled to a nongravitational $(d+1)$-dimensional bath. This coupling is achieved by imposing transparent boundary conditions for some matter fields in the AdS$_{d+1}$ near its asymptotic boundary. An explicit island model can be constructed as follows. We consider a gravitational AdS$_{d+1}$ spacetime with a free massive scalar field  $\phi_{1}(x,z)$ as the transparent matter field. We model the nongravitational bath by another AdS$_{d+1}$ spacetime with a matter field $\phi_{2}(x,z)$ which is also a free massive scalar field and with the same mass as $\phi_{1}(x,z)$.\footnote{More precisely, in the nongravitational AdS the field $\phi_{2}(x,z)$ has the same $ml_{AdS}$ as the field $\phi_{1}(x,z)$ in the gravitational AdS$_{d+1}$. We will set the all AdS length scales to one hereafter.} These two AdS$_{d+1}$'s are glued together along their asymptotic boundary such that the energy of $\phi_{1}(x,z)$ in the gravitational AdS$_{d+1}$ can freely leak into $\phi_{2}(x,z)$ in the nongravitational AdS$_{d+1}$ (see Fig.\ref{pic:islandmodel}). This leaky boundary condition is achieved by identifying the asymptotic behaviors of the two fields $\phi_{1}(x,z)$ and $\phi_{2}(x,z)$ as
\begin{equation}
    \begin{split}
        \phi_{1}(x,z)&=\Big(\alpha(x)z^{\Delta}+\mathcal{O}(z^{\Delta+2})\Big)+\Big(\frac{1}{2\Delta-d}\beta(x)z^{d-\Delta}+\mathcal{O}(z^{d-\Delta+2})\Big)\,,\\\phi_{2}(x,z)&=\Big(\beta(x)z^{d-\Delta}+\mathcal{O}(z^{d-\Delta+2})\Big)+\Big(\frac{1}{d-2\Delta}\alpha(x)z^{\Delta}+\mathcal{O}(z^{\Delta+2})\Big)\,,\label{eq:asympcouple}
    \end{split}
\end{equation}
where $\alpha(x)$ and $\beta(x)$ are both dynamical modes, $\Delta=\frac{d}{2}+\sqrt{\frac{d^2}{4}+m^{2}}$ with $m^2$ the mass square of the two scalar fields and the above expansion is done in the AdS$_{d+1}$ Poincar\'{e} patch near the asymptotic boundary $z=0$. The metric in the Poincar\'{e} patch is
\begin{equation}
ds^2=\frac{dz^{2}+\eta_{ij}dx^{i}dx^{j}}{z^{2}}\,,\label{eq:Poincare}
\end{equation}
where $\eta_{ij}$ is the metric of flat $d$-dimensional Minkowski spacetime.

A comprehensive way to understand the above model is to go to its dual description. One can dualize the gravitational AdS$_{d+1}$ to a CFT$_{d}$ that lives on its asymptotic boundary $z=0$ using the AdS/CFT correspondence \cite{Maldacena:1997re,Gubser:1998bc,Witten:1998qj}. Thus the dual description of the above island model is given by a CFT$_{d}$ coupled to a $(d+1)$-dimensional nongravitational bath. The geometry of the bath is given by Equ.~(\ref{eq:Poincare}) and the CFT$_{d}$ lives on its asymptotic boundary $z=0$. The coupling is achieved by a double-trace deformation, for which the full Hamiltonian is given by
\begin{equation}
    H_{\text{tot}}=H_{\text{CFT$_{d}$}}+g\int d^{d}x \mathcal{O}_{1}(x)\mathcal{O}_{2}(x)+H_{\text{bath}}\,,\label{eq:coupledH}
\end{equation}
where $\mathcal{O}_{1}(x)$ is the CFT$_{d}$ single-trace operator that is dual to the bulk field $\phi_{1}(x,z)$, $\mathcal{O}_{2}(x)$ is the boundary extrapolation of the bath field $\phi_{2}(x,z)$ and $g$ is the coupling constant. The extrapolation is given by
\begin{equation}
    \phi_{2}(x,z)\rightarrow \mathcal{O}_{2}(x)z^{d-\Delta}\,,\quad \text{as $z\rightarrow0$}\,.
\end{equation}
The dimension of the operator $\mathcal{O}_{1}(x)$ is $\Delta_{1}=\Delta=\frac{d}{2}+\sqrt{\frac{d^2}{4}+m^2}$ and that of $\mathcal{O}_{2}(x)$ is $\Delta_{2}=d-\Delta$ so from the CFT$_{d}$ perspective the coupling in Equ.~(\ref{eq:coupledH}) is a marginal deformation. From Equ.~(\ref{eq:coupledH}) one can easily work out Equ.~(\ref{eq:asympcouple}) following the proposal in \cite{Witten:2001ua} for double-trace deformation in AdS/CFT.\footnote{See \cite{Geng:2023ynk} for a derivation and refinement of the proposal in \cite{Witten:2001ua} and some worked-out examples including Equ.~(\ref{eq:asympcouple}).} Using Equ.~(\ref{eq:coupledH}) and conformal perturbation theory one can work out the time-ordered Green function of the scalar field $\phi_{1}(x,z)$ in the gravitational AdS$_{d+1}$ as
    \begin{equation}
\begin{split}
\langle \mathbf{T}\phi_{1}(x_{1},z_{1})\phi_{1}(x_{2},z_{2})\rangle&=a_{1}2^{-\Delta_{1}}\frac{\Gamma[\Delta_{1}]}{\pi^{\frac{d}{2}}(2\Delta_{1}-d)\Gamma[\Delta_{1}-\frac{d}{2}]}\frac{_{2}F_{1}[\frac{\Delta_{1}}{2},\frac{\Delta_{1}+1}{2},\Delta_{1}-\frac{d}{2}+1,\frac{1}{Z^{2}}]}{(Z)^{\Delta_{1}}}\\&+a_{2}2^{-\Delta_{2}}\frac{\Gamma[\Delta_{2}]}{\pi^{\frac{d}{2}}(2\Delta_{2}-d)\Gamma[\Delta_{2}-\frac{d}{2}]}\frac{_{2}F_{1}[\frac{\Delta_{2}}{2},\frac{\Delta_{2}+1}{2},\Delta_{2}-\frac{d}{2}+1,\frac{1}{Z^{2}}]}{(Z)^{\Delta_{2}}},
\end{split}
\end{equation}
where $Z=\frac{z_{1}^{2}+z_{2}^{2}+\eta_{ij}(x_{1}-x_{2})^{i}(x_{1}-x_{2})^{j}}{2z_{1}z_{2}}$ is the invariant distance between the two operator insertions in AdS$_{d+1}$.\footnote{For simplicity, we have ignored the $i\epsilon$-prescription.} According to Equ.~(\ref{eq:asympcouple}), we have, to the leading order in the coupling constant, the relation
\begin{equation}
    a_{2}=\frac{g^{2}}{(2\Delta-d)^{2}}a_{1}\,.
\end{equation}
A caveat is that in the above model both $\Delta_{1}$ and $\Delta_{2}$ should be above the CFT$_{d}$ unitarity bound $\frac{d-2}{2}$. As we will see in the next section, this is important to avoid potential instabilities.

\subsection{The Island Model as a Higgs Phase}\label{sec:higgs}
It was shown by relevant calculations and various compelling indirect arguments in the early works \cite{Porrati:2001gx,Porrati:2002dt,Porrati:2003sa,Duff:2004wh,Aharony:2006hz} that the graviton in the island model is in fact massive and the mass is generated by the spontaneous breaking of the diffeomorphism symmetry due to the bath coupling. The recent progress made in \cite{Geng:2023ynk,Geng:2023zhq} provided a fully covariant path integral derivation of the above results. Given the importance of these results to the later discussions, let's concisely review them in this subsection.

Let's focus on the gravitational AdS$_{d+1}$ in the island model and work to the leading order in the Newton's constant $G_{N}$. In this subsection, we will use $x^{\mu}$ to collectively denote the AdS$_{d+1}$ coordinates $(x^{i},z)$. We have the path integral
\begin{equation}
\begin{split}
Z_{H}&=\int D[\psi]D[h_{\mu\nu}]e^{iS_{\text{matter}}[\psi;g^{0}]+iS[h;g^{0}]+iS_{\text{int}}[\psi,h;g^{0}]}\,,\\S_{\text{int}}[\psi,h;g^{0}]&=\frac{\sqrt{16\pi G_{N}}}{2}\int d^{d+1}x\sqrt{-g^{0}(x)}h_{\mu\nu}(x)T^{\mu\nu}_{\text{matter}}(x)\,,\label{eq:Z}
\end{split}
\end{equation}
where we use $\psi$ to collectively denote matter fields including the transparent matter field $\phi_{1}$, $g^{0}_{\mu\nu}$ denotes the background geometry Equ.~(\ref{eq:Poincare}), $\sqrt{16\pi G_{N}}h_{\mu\nu}$ denotes the fluctuation of the background metric and $S[h;g^{0}]$ denotes the kinetic term of the metric fluctuation, i.e. the graviton field. At this leading order of $G_{N}$, the diffeomorphism transformation is given by
\begin{equation}
\begin{split}
&x^{\mu}\rightarrow x^{\mu}, \quad g^{0}_{\mu\nu}(x)\rightarrow g^{0\prime}_{\mu\nu}(x)=\frac{\partial x'^{\rho}}{\partial x^{\mu}}\frac{\partial x'^{\sigma}}{\partial x^{\nu}}g^{0}_{\rho\sigma}(x')\,,\quad\psi(x)\rightarrow\psi'(x)=\psi(x')\,,\\&h_{\mu\nu}(x)\rightarrow h_{\mu\nu}'(x)=\frac{\partial x'^{\rho}}{\partial x^{\mu}}\frac{\partial x'^{\sigma}}{\partial x^{\nu}}h_{\rho\sigma}(x')+\nabla_{\mu}\epsilon_{\nu}(x)+\nabla_{\nu}\epsilon_{\mu}(x)\,,\quad \text{where } x'^{\mu}=x^{\mu}+\sqrt{16\pi G_{N}}\epsilon^{\mu}(x)\,,\label{eq:diffeotr}
\end{split}
\end{equation}
and the total action $S_{\text{tot}}[\psi,h;g^{0}]=S_{\text{matter}}[\psi;g^{0}]+S[h;g^{0}]+S_{\text{int}}[\psi,h;g^{0}]$ in Equ.~(\ref{eq:Z}) transforms as
\begin{equation}
    S_{\text{tot}}[\psi,h;g^{0}]\rightarrow  S_{\text{tot}}[\psi,h;g^{0}]+\frac{\sqrt{16\pi G_{N}}}{2}\int d^{d+1}x\sqrt{-g^{0}(x)}\nabla_{\mu}\epsilon_{\nu}(x) T^{\mu\nu}_{\text{matter}}(x)\,,
\end{equation}
where we have used the fact that $T^{\mu\nu}_{\text{matter}}(x)$ is a symmetric tensor. In the standard AdS/CFT context \cite{Aharony:1999ti}, one imposes the reflective boundary condition for matter fields so one would have
\begin{equation}
\begin{split}
    &\int d^{d+1}x\sqrt{-g^{0}(x)}\nabla_{\mu}\epsilon_{\nu}(x) T^{\mu\nu}_{\text{matter}}(x)\,,\\&=\int d^{d+1}x\sqrt{-g^{0}(x)}\nabla_{\mu}\Big(\epsilon_{\nu}(x) T^{\mu\nu}_{\text{matter}}(x)\Big)-\int d^{d+1}x\sqrt{-g^{0}(x)}\epsilon_{\nu}(x) \nabla_{\mu}T^{\mu\nu}_{\text{matter}}(x)\,,\\&=0\,,
    \end{split}
\end{equation}
where in the second line the first term is a total derivative term and it is zero due to the reflective boundary condition, i.e. the vanishing of the boundary flux, and the second term is zero due to the local conservation of the stress-energy tensor everywhere in the gravitational AdS$_{d+1}$. However, none of the above features persist in the island model where there exist transparent matter fields in the gravitational AdS$_{d+1}$. Hence, to have a fully covariant, i.e. diffeomorphism invariant, description, we can introduce a vector field $V^{\mu}(x)$ which transforms under the diffeomorphism Equ.~(\ref{eq:diffeotr}) as
\begin{equation}
    V_{\mu}(x)\rightarrow V'_{\mu}(x)=\frac{\partial x^{\nu\prime}}{\partial x^{\mu}} V_{\nu}(x')-\epsilon_{\mu}(x)\,,\label{eq:diffV}
\end{equation}
and integrate it into the path integral Equ.~(\ref{eq:Z}) as
\begin{equation}
    Z_{\text{full}}=\int D[V^{\mu}]D[\psi]D[h_{\mu\nu}]e^{iS_{\text{matter}}[\psi;g^{0}]+iS[h;g^{0}]+iS_{\text{int}}[\psi,h;g^{0}]+i\sqrt{16\pi G_{N}}\int d^{d+1}x\sqrt{-g^{0}(x)}\nabla_{\mu}V_{\nu}(x)T^{\mu\nu}_{\text{matter}}(x)}\,.\label{eq:Zfull}
\end{equation}
This full partition function is invariant under the diffeomorphism transform Equ.~(\ref{eq:diffeotr}) together with Equ.~(\ref{eq:diffV}) to leading order in $G_{N}$. Moreover, Equ.~(\ref{eq:Z}) can be realized as a gauge-fixed version of Equ.~(\ref{eq:Zfull}) by choosing the unitary gauge $V_{\mu}(x)=0$. At this point, one should realize that this is similar to the standard story in the Higgs mechanism where the unitary gauge fixes the Goldstone boson to zero and the Goldstone boson disappears in the path integral, i.e. swallowed by the massive gauge boson \cite{peskin1995iqf}. Nevertheless, to prove that this is indeed the case one has to show that there is a St\"{u}ckelberg mass term for the graviton with $V^{\mu}(x)$ as the St\"{u}ckelberg/Goldstone vector field. This can be done by integrating out the transparent matter field $\phi_{1}(x)$ to get the effective action for the graviton and the vector field $V^{\mu}(x)$. This effective action is invariant under the transform Equ.~(\ref{eq:diffeotr}) together with Equ.~(\ref{eq:diffV}) so it is only a functional of the invariant combination
\begin{equation}
    h_{\mu\nu}(x)+\nabla_{\mu}V_{\nu}(x)+\nabla_{\nu}V_{\mu}(x)\,.\label{eq:hV}
\end{equation}
We are interested in the quadratic terms of this invariant combination in the low energy effective action. This effective action can be obtained by firstly sending the graviton field to zero, i.e. only considering the interaction between the transparent matter field and the vector field $V^{\mu}(x)$, and then replacing $\nabla_{\mu}V_{\nu}(x)+\nabla_{\nu}V_{\mu}(x)$ in the resulting effective action with the invariant combination Equ.~(\ref{eq:hV}) \cite{Geng:2023ynk,Geng:2023zhq}. Hence, we want to firstly compute
\begin{equation}
\begin{split}
    \delta S_{2}[V]=i\frac{1}{2}(\sqrt{16\pi G_{N}})^{2}\int d^{d+1}xd^{d+1}y\sqrt{-g^{0}(x)}\sqrt{-g^{0}(y)}\nabla_{\mu}V_{\nu}(x)\nabla_{\rho}V_{\sigma}(y)\langle \mathbf{T} T^{\mu\nu}(x) T^{\rho\sigma}(y)\rangle\,,\label{eq:Vbare}
\end{split}
\end{equation}
where $T^{\mu\nu}(x)$ is the stress-energy tensor of the transparent matter field $\phi_{1}(x)$ and the low energy behavior is controlled by the large distance behavior of the two-point function 
\begin{equation}
    \Pi^{\mu\nu,\rho\sigma}(x,y)=\langle \mathbf{T} T^{\mu\nu}(x) T^{\rho\sigma}(y)\rangle\,.
\end{equation}
It is nicely computed in \cite{Aharony:2006hz} that the large distance limit of this two point function is 
\begin{equation}
    \begin{split}
        16\pi G_{N}\Pi^{\mu\nu,\rho\sigma}(x,y)=4M^{2}\nabla^{(\mu}D^{\nu),(\sigma}(x,y)\nabla^{\rho)}\,,\label{eq:TT}
        \end{split}
\end{equation}
where $D^{\nu,\sigma}(x,y)$ is in the form of the propagator of a massive vector field satisfying\footnote{Thus, the mass square of the vector is $m^{2}_{V}=2d$.}
\begin{equation}
(\nabla^{2}-d)D^{\nu,\sigma}(x,y)=i\frac{g^{0\nu\sigma}}{\sqrt{-g}}\delta^{d+1}(x-y)\,,\label{eq:Green}
\end{equation}
and $M^{2}$ is given in general in \cite{Aharony:2006hz} as
\begin{equation}
M^{2}=-G_{N}\frac{2^{4-d}\pi^{\frac{3-d}{2}}}{(d+2)\Gamma(\frac{d+3}{2})}\frac{a_{1}a_{2}\Delta_{1}\Delta_{2}\Gamma[\Delta_{1}]\Gamma[\Delta_{2}]}{\Gamma[\Delta_{1}-\frac{d}{2}]\Gamma[\Delta_{2}-\frac{d}{2}]}\,.\label{eq:mass}
\end{equation}
We note that massive vector field is automatically divergenceless on-shell so the propagator $D^{\nu,\sigma}(x,y)$ is also divergenceless. Using the above results and integrating $x$ by part in Equ.~(\ref{eq:Vbare}) we have
\begin{equation}
    \begin{split}
    \delta S_{2}[V]&=-i\frac{1}{2}(\sqrt{16\pi G_{N}})^{2}\int d^{d+1}xd^{d+1}y\sqrt{-g^{0}(x)}\sqrt{-g^{0}(y)}V_{\nu}(x)\nabla_{\rho}V_{\sigma}(y)\nabla_{\mu}\langle \mathbf{T} T^{\mu\nu}(x) T^{\rho\sigma}(y)\rangle\,,\\&=\frac{1}{2}2M^{2}\int d^{d+1}x\sqrt{-g^{0}(x)}V_{\nu}(x)g^{0\nu(\sigma}(x)\nabla^{\rho)}\nabla_{\rho}V_{\sigma}(x)\,,\\&=\frac{M^{2}}{2}\int d^{d+1}x\sqrt{-g^{0}(x)}\Big[V_{\nu}(x)\nabla^{\mu}\nabla_{\mu}V^{\nu}(x)+V_{\nu}(x)\nabla^{\mu}\nabla^{\nu}V_{\mu}(x)\Big]\,,\\&=-\frac{M^{2}}{2}\int d^{d+1}x\sqrt{-g^{0}(x)}\Big[\nabla^{\mu}V^{\nu}(x)\nabla_{\mu}V_{\nu}(x)+\nabla^{\mu}V^{\nu}(x)\nabla_{\nu}V_{\mu}(x)\Big]\,,\\&=-\frac{M^{2}}{4}\int d^{d+1}x\sqrt{-g^{0}(x)}\Big(\nabla^{\mu}V^{\nu}(x)+\nabla^{\nu}V^{\mu}(x)\Big)\Big(\nabla_{\mu}V_{\nu}(x)+\nabla_{\nu}V_{\mu}(x)\Big)\,,\label{eq:Valmost}
\end{split}
\end{equation}
which is valid only in the large distance limit. Due to the fact that $D^{\nu,\sigma}(x,y)$ is divergenceless, the two point function Equ.~(\ref{eq:TT}) is traceless in both $\mu\nu$ and $\rho\sigma$. Therefore, the divergence of $V^{\mu}(x)$, i.e. $\nabla_{\mu}V^{\mu}(x)$, decouples so in the correct $\delta S_{2}[V]$ the divergence part of $V_{\nu}(x)$ has to decouple. In other words, the propagator of $V^{\mu}(x)$ has to be divergenceless. The following action satisfies the above requirement
\begin{equation}
    \delta S_{2}[V]=-\frac{M^{2}}{4}\int d^{d+1}x\sqrt{-g^{0}(x)}\Big[\Big(\nabla^{\mu}V^{\nu}(x)+\nabla^{\nu}V^{\mu}(x)\Big)\Big(\nabla_{\mu}V_{\nu}(x)+\nabla_{\nu}V_{\mu}(x)\Big)-4\nabla^{\mu}V_{\mu}(x)\nabla^{\nu}V_{\nu}(x)\Big]\,.\label{eq:Vfinal}
\end{equation} 
This can be seen by first working out the equation of motion of the vector field obtained from the action Equ.~(\ref{eq:Vfinal})
\begin{equation}
    \nabla^{2}V^{\nu}(x)+2\nabla_{\mu}\nabla^{\nu}V^{\mu}(x)-2\nabla^{\nu}\nabla_{\mu}V^{\mu}(x)=0\,,
\end{equation}
which is equivalent to
\begin{equation}
    \nabla_{\mu}(\nabla^{\mu}V^{\nu}-\nabla^{\nu}V^{\mu})(x)-2d V^{\nu}(x)=0\,,
\end{equation}
where we used $R_{\mu\nu}=-dg_{\mu\nu}$ for AdS$_{d+1}$, and this is exactly the equation of motion of a massive vector field. Then it is easy to see that $V^{\nu}(x)$ is divergenceless using this equation of motion. Now we can restore the $h_{\mu\nu}(x)$ dependence by the replacement to the invariant combination Equ.~(\ref{eq:hV}) which gives
\begin{equation}
    S_{\text{eff}}^{(2)}[h,V]=-\frac{M^{2}}{4}\int d^{d+1}x\sqrt{-g^{0}(x)}\Big(\tilde{h}_{\mu\nu}(x)\tilde{h}^{\mu\nu}(x)-\tilde{h}^{2}(x)\Big)\,,\label{eq:stuckelberg}
\end{equation}
where $\tilde{h}_{\mu\nu}(x)$ denotes the invariant combination Equ.~(\ref{eq:hV}) and $\tilde{h}(x)$ is the trace of $\tilde{h}_{\mu\nu}(x)$ under the background metric $g^{0}_{\mu\nu}$. 

As a result, we indeed get a St\"{u}ckelberg mass term for the graviton with $V^{\mu}(x)$ as the St\"{u}ckelberg/Goldstone vector field. We should realize that this term is indeed generated by a Higgs mechanism as a one-loop quantum effect and one can think of the Higgs field as the transparent matter field $\phi_{1}(x)$ (see Sec.~\ref{sec:mechanism} for more details). Furthermore, the fact that not only $\Delta_{1}$ but also $\Delta_{2}$ is above the unitarity bound $\frac{d-2}{2}$ ensures that the mass square of the graviton in Equ.~(\ref{eq:mass}) is positive and so no instability is generated. Interestingly, if we choose the unitary gauge, i.e. setting $V^{\mu}(x)=0$, we will see that the graviton mass term Equ.~(\ref{eq:stuckelberg}) becomes the famous Feirz-Pauli mass term of the graviton \cite{Hinterbichler:2011tt}. This completes the proof that the island model is in fact in the Higgs phase where the graviton is massive due to the spontaneously broken diffeomorphism symmetry.

\subsection{Holography in the Island Model}\label{sec:holography}
It was pointed out by \cite{Geng:2021hlu,Geng:2023zhq} that the holographic interpretation in the island model that the island region $\mathcal{I}$ in the gravitational AdS$_{d+1}$ is part of the entanglement wedge of the bath subregion $R$ is inconsistent with the long-range property of the standard massless gravity. This is based on the observation that in the standard massless gravity the gravitational Gauss' law asserts that the Hamiltonian of the gravitational AdS$_{d+1}$ in fact is a boundary term and therefore is fully accessible near its asymptotic boundary, i.e. purely in the complement of the island region. Hence if island exists, one would have an operator in $\bar{\mathcal{I}}$, i.e. the Hamiltonian, that does not commute with operators inside the island region $\mathcal{I}$, which contradicts the above holographic interpretation of the island. This is because $\bar{\mathcal{I}}$ is part of the entanglement wedge of $\bar{R}$ and so operators in $\mathcal{I}$ and operators in $\bar{\mathcal{I}}$ should exactly commute with each other as operators in $R$ and operators in $\bar{R}$.

One can formulate the above intuition in a precise way in general relativity using the ADM formalism \cite{Arnowitt:1962hi}. The ADM formalism starts with the ADM decomposition of the metric \cite{Arnowitt:1962hi} on a (d+1)-dimensional spacetime
\begin{equation}
ds^{2}=-N^{2} dt^{2}+g_{ij}(dx^{i}+N^{i}dt)(dx^{j}+N^{j}dt)\,,\label{eq:ADM}
\end{equation}
where N is called the \textit{lapse function}, the vector $N^{i}$ is called the \textit{shift vector}. This decomposition can be thought of as a gauge fixing procedure which fixes the $(d+1)$-coordinate reparametrizations. One can use the Gauss-Codazzi equation to decompose the matter-coupled Einstein-Hilbert action as
\begin{equation}
S=\frac{1}{16\pi G_{N}}\int dt d^{d}x N\sqrt{g}(R[g]-2\Lambda+K_{ij}K^{ij}-K^{2})+S_{\text{matter}}+S_{\text{bdy}}\,,\label{eq:actionADM}
\end{equation}
where $\Lambda=-\frac{d(d-1)}{2}$ is the cosmological constant, $R[g]$ is the Ricci scalar of the spatial metric $g_{ij}$, $S_{\text{matter}}$ is the matter field action and $K_{ij}$ denotes the extrinsic curvature of the hypersurfaces with constant $t$-coordinate. The boundary term $S_{\text{bdy}}$ in the action ensures a well-defined variational principle of this action. The extrinsic curvature  of the constant-$t$ hypersurfaces can be calculated using the metric Equ.~(\ref{eq:ADM}) as
\begin{equation}
K_{ij}=\frac{1}{2N}(-\dot{g}_{ij}+D_{j}N_{i}+D_{i}N_{j})\,,\label{eq:K}
\end{equation}
where $D_{i}$ is the torsion-free and metric-compatible covariant derivative with respect to the spatial metric $g_{ij}$. From Equ.~(\ref{eq:actionADM}) and Equ.~(\ref{eq:K}), we can see that the canonical momenta associated with the lapse function $N$ and shift vector $N^{i}$ are zero
\begin{equation}
\Pi=\frac{1}{\sqrt{g}}\frac{\delta S}{\delta \dot{N}}=0\,,\quad \Pi_{i}=\frac{1}{\sqrt{g}}\frac{\delta S}{\delta \dot{N^{i}}}=0\,.\label{eq:primary}
\end{equation}
Furthermore, the Hamiltonian of the system described by Equ.~(\ref{eq:actionADM}) can be written as
\begin{equation}
H_{\text{tot}}=\int d^{d}x\sqrt{g}\Big[N\mathcal{H}+N^{i}\mathcal{H}_{i}\Big]+H_{\text{bdy}}\,,\label{eq:Htot}
\end{equation}
where
\begin{equation}
\begin{split}
  \mathcal{H}&=16\pi G_{N}\Big(\Pi_{ij}\Pi^{ij}-\frac{1}{d-1}(\Pi^{i}_{i})^{2}\Big)-\frac{1}{16\pi G_{N}}(R[g]-2\Lambda)+\mathcal{H}_{\text{matter}}\,,\\
\mathcal{H}_{i}&=-2g_{ij}D_{k}\Pi^{jk}+\mathcal{H}_{i,\text{matter}}\,.\label{eq:constraints}
  \end{split}
\end{equation}
In the above equations, $\mathcal{H}_{\text{matter}}$ is the Hamiltonian density of the matter field and $\mathcal{H}_{i,\text{matter}}$ is the momentum density of the matter fields with $\Pi^{ij}$ as the canonical momentum of the spatial metric $g_{ij}$, i.e.
\begin{equation}
\Pi^{ij}=\frac{1}{\sqrt{g}}\frac{\delta S}{\delta \dot{g}_{ij}}=-\frac{1}{16\pi G_{N}}\Big(K^{ij}-g^{ij}K\Big)\,.
\end{equation}
As a constrained system, the equations Equ.~(\ref{eq:primary}) are called \textit{primary constraints} \cite{dirac2001quantum}. The primary constraints have to be preserved under the time evolution generated by the Hamiltonian Eq.~(\ref{eq:Htot}) and this generates the following \textit{secondary constraints}
\begin{equation}
\mathcal{H}=0\,,\quad \mathcal{H}_{i}=0\,.
\end{equation}
In quantum theory, secondary constraints constrain the physical states and observables once we promote $\mathcal{H}$ and $\mathcal{H}_{i}$ to operators $\hat{\mathcal{H}}$ and $\hat{\mathcal{H}}_{i}$. More precisely, the physical states in the Hilbert space of the system have to be annihilated by $\hat{\mathcal{H}}$ and $\hat{\mathcal{H}}_{i}$\footnote{See however \cite{Giddings:2022hba} for the possible relax.} and gauge invariant observables have to commute with them. We are mostly interested in exploiting the Hamiltonian constraint $\mathcal{H}=0$ as the momentum constraints $\mathcal{H}_{i}=0$ basically require the diffeomorphism invariance in the spatial directions and are easily satisfied.

In our context, we are working around the AdS background Equ.~(\ref{eq:metric}) and we can analyze the Hamiltonian constraint perturbatively around this background metric in which $N=\frac{1}{z}$, $N_{i}=0$ and the spatial background metric is $g^{0}_{ij}=\frac{1}{z^{2}}\delta_{ij}$. Since the lapse function and the shift vector are constrained to have a trivial dynamics from Equ.~(\ref{eq:primary}), we only have to focus on the dynamics of the spatial metric $g_{ij}$. For our purpose, the perturbative dynamics to the leading nontrivial order in $G_{N}$ is enough. Therefore we will treat the metric $g_{ij}$ perturbatively and expand it around the background as
\begin{equation}
g_{ij}=g_{ij}^{0}+\sqrt{16\pi G_{N}}h_{ij}\,,
\end{equation}
where similar to the discussions in Sec.~\ref{sec:higgs}, we can think of $h_{ij}$ as the graviton field which has a normalized kinetic term from expanding the Einstein-Hilbert action to quadratic order in $h_{ij}$. To the leading order in $G_{N}$, we have the zeroth order Hamiltonian constraint
\begin{equation}
\mathcal{H}^{(0)}=-\frac{1}{16\pi G_{N}}(R[g^{0}]-2\Lambda)=0\,,
\end{equation}
which is constantly satisfied using the background geometry $g^{0}_{ij}=\frac{1}{z^{2}}\delta_{ij}$ and $\Lambda=-\frac{d(d-1)}{2}$. Thus the nontrivial dynamics from the Hamiltonian constraint starts from the first order in $G_{N}$, where we have
\begin{equation}
\mathcal{H}^{(1)}=-\frac{1}{\sqrt{16\pi G_{N}}}\Big[(d-1)h+\hat{\nabla}_{i}\hat{\nabla}_{j}h^{ij}-\hat{\nabla}^{2}h\Big]+\mathcal{H}_{\text{matter}}+\mathcal{H}_{\text{graviton}}=0\,,\label{eq:core}
\end{equation}
in which $h$ is the trace of $h_{ij}$ under the background metric $g^{0}_{ij}$ and $\hat{\nabla}_{i}$ is the torsion free and metric compatible covariant derivative with respect to the same background metric. In the above equation, $\mathcal{H}_{\text{matter}}$ is the Hamiltonian density of matter fields in the background geometry Equ.~(\ref{eq:metric}) and $\mathcal{H}_{\text{graviton}}$ is the free graviton Hamiltonian in the same background geometry. For the sake of simplicity, let's first focus on the matter source and ignore $\mathcal{H}_{\text{graviton}}$ for a moment. Using Equ.~(\ref{eq:core}), we can see that the matter source and the metric fluctuation satisfy
\begin{equation}
\begin{split}
\sqrt{16\pi G_{N}}H_{\text{matter}}&=\sqrt{16\pi G_{N}}\int d^{d}x
\sqrt{g^{0}}N\mathcal{H}_{\text{matter}}=\int d^{d}x
\sqrt{g^{0}}N\Big[(d-1)h+\hat{\nabla}_{i}\hat{\nabla}_{j}h^{ij}-\hat{\nabla}^{2}h\Big]\,,\\&=\int d^{d}x
\sqrt{g^{0}}\hat{\nabla}_{i}\Big[\frac{\hat{\nabla}_{j}h^{ij}-\hat{\nabla}^{i}h}{z}+\frac{1}{z^{2}}h^{zi}-\frac{1}{z^{2}}h\delta^{i}_{z}\Big]\,,\\&=\int d^{d}x\sqrt{g^{0}}\hat{\nabla}_{i}\Bigg[N\Big[\hat{\nabla}_{j}h^{ij}-\hat{\nabla}^{i}h+\frac{1}{z} (h^{zi}-h\delta^{i}_{z} )\Big]\Bigg]\,,\\&\equiv H_{\partial}\,,\label{eq:HADM}
\end{split}
\end{equation}
which is thus a boundary term. In our case, as standard in the study of holography, we imposes the Dirichlet boundary condition for the bulk metric fluctuation or the graviton field near the asymptotic boundary of the gravitational AdS. Thus the last two terms in Equ.~(\ref{eq:HADM}) are zero. As a result, one can see that the right hand side of Equ.~(\ref{eq:HADM}) is now exactly given by the ADM energy $\sqrt{16\pi G_{N}}\hat{H}_{\text{ADM}}$ for asymptotic AdS spacetimes \cite{Arnowitt:1962hi,Hawking:1995fd,Giddings:2018umg}. This result has a rather remarkable implication in quantum gravity, as physical (gauge invariant) operators $\hat{O}(x)$ should commute with the constraint $\hat{\mathcal{H}}$. The integrated version of this constraint to the first nontrivial order in $G_{N}$ is
\begin{equation}
    [\sqrt{16\pi G_{N}}\hat{H}_{\text{matter}}-\hat{H}_{\partial},\hat{O}(x)]=0\,.\label{eq:commutator}
\end{equation}
This implies that
\begin{equation}
[\hat{H}_{\partial},\hat{O}(x)]=\sqrt{16\pi G_{N}}[\hat{H}_{\text{matter}},\hat{O}(x)]=-i\sqrt{16\pi G_{N}} \frac{\partial}{\partial t}\hat{O}(x)\,,\label{eq:constraintO}
\end{equation}
which is in fact true to all orders in $G_{N}$ \cite{Chowdhury:2021nxw}. Explicit solutions of Equ.~(\ref{eq:constraintO}) can be constructed using gravitational Wilson lines \cite{Donnelly:2018nbv,Giddings:2018umg} (see also Sec.~\ref{sec:gravKR}). In summary, we can see that in the standard massless gravity there exists an operator $\hat{H}_{\partial}$ which is accessible near the asymptotic boundary of AdS$_{d+1}$ that doesn't commute with any operator in the bulk. This is a genuinely gravitational effect as it is controlled by the nonvanishingness of $G_{N}$. As we have discussed, this is in contradiction with the potential existence of island as a closed nonempty subregion in the gravitational AdS$_{d+1}$ for the standard massless gravity. One should note that this is not a surprising fact as we expect that in the standard massless gravity we have the usual notion of holography that information in the gravitational bulk is fully encoded in its asymptotic boundary.

Interestingly, the above issue is nicely avoided in the island model due to the fact that in the island model the graviton is massive \cite{Geng:2021hlu,Geng:2023zhq,Geng:2025tba1}. More precisely, this is because of the additional graviton mass term Equ.~(\ref{eq:stuckelberg}) which will modify the Hamiltonian constraint Equ.~(\ref{eq:constraints}). The modification can be easily figured out by noticing that the Hamiltonian constraint is in fact the $00$-component of the matter coupled Einstein's equation in the ADM decomposition Equ.~(\ref{eq:ADM}). As before, we will perturbatively expand the modified constraint to the first nontrivial order in $G_{N}$ which is now
\begin{equation}
\begin{split}
    \mathcal{H}^{(1)}=-\frac{1}{\sqrt{16\pi G_{N}}}\Big[(d-1)h+\hat{\nabla}_{i}\hat{\nabla}_{j}h^{ij}-\hat{\nabla}^{2}h\Big]+\mathcal{H}_{\text{matter}}+\mathcal{H}_{\text{graviton}}-\frac{M^{2}}{\sqrt{16\pi G_{N}}}\tilde{h}_{i}^{i}(x)=0\,,\label{eq:stuckelbergH}
    \end{split}
\end{equation}
where $\tilde{h}_{i}^{i}$ is given by the invariant combination
\begin{equation}
    \tilde{h}_{i}^{i}=h_{i}^{i}+2\nabla_{i}V^{i}\,,
\end{equation}
where $\nabla_{\mu}$ is the torsion free and metric compatible covariant derivative with the full background metric Equ.~(\ref{eq:metric}). We in fact have the conjugate momentum of the $0$-th component $V^{0}(x)$ of the St\"{u}ckelberg vector field $V^{\mu}(x)$
\begin{equation}
    \hat{\pi}_{V^{0}}(x)=M^{2}\Big(h_{i}^{i}+2\nabla_{i}\hat{V}^{i}\Big)(x)\,,
\end{equation}
for which we have the equal-time commutator
\begin{equation}
    [\hat{\pi}_{V^{0}}(\vec{x},t),\hat{V}^{0}(\vec{y},t)]=-i\frac{1}{N\sqrt{g^{0}}(x)}\delta^{d}(\vec{x}-\vec{y})\,.
\end{equation}
As a result, we have the integrated constraint
\begin{equation}
    \sqrt{16\pi G_{N}}\Big(\hat{H}_{\text{matter}}+\hat{H}_{\text{graviton}}\Big)-\hat{\Pi}_{V^{0}}+\hat{H}_{\partial}=0\,,\label{eq:constraintMass}
\end{equation}
which constrains physical operators and we have defined 
\begin{equation}
    \hat{\Pi}_{V^{0}}=\int d^{d}\vec{x} N\sqrt{g^{0}}\hat{\pi}_{V^{0}}(x,t)\,.
\end{equation}
It is easy to construct operators that satisfy the constraint Equ.~(\ref{eq:constraintMass}) and commute with the ADM Hamiltonian. An explicit example is the following. Given a scalar matter field operator $\hat{O}(x)$, we have the physical operator
\begin{equation}
    \hat{O}^{\text{Phys}}(x)=\hat{O}(x+\sqrt{16\pi G_{N}}\hat{V}(x))\,,\label{eq:Ophys}
\end{equation}
i.e. we replace the coordinate $x^{\mu}$ by the invariant combination $x^{\mu}+\hat{V}^{\mu}(x)$.\footnote{We should notice that the dressed operator $\hat{O}(x+\sqrt{16\pi G_{N}}\hat{V}(x))$ should really by understood using the Taylor expansion of the operator $\hat{O}$ around $x$. Each term in the Taylor expansion is a local operator at $x$ and higher order terms in the expansion are suppressed by higher powers of $\sqrt{G_{N}}$. Thus, in the semiclassical limit with $G_{N}\ll 1$, this dressed operator is sensibly localized at $x$.} It is easy to check that we have
\begin{equation}
[\sqrt{16\pi G_{N}}\Big(\hat{H}_{\text{matter}}+\hat{H}_{\text{graviton}}\Big)-\hat{\Pi}_{V^{0}},\hat{\mathcal{O}}^{\text{Phys}}(x)]=0\,,\quad\text{and }[\hat{H}_{\partial},\hat{\mathcal{O}}^{\text{Phys}}(x)]=0\,.
\end{equation}
 In summary, we can see that the fact that the island model is in the Higgs phase with a massive graviton enables us to have operators in the gravitational AdS$_{d+1}$ that commutes with the ADM Hamiltonian. This avoids the tension between the existence of island and holography in the standard massless gravity theories. The Stu\"{u}ckelberg/Goldstone vector field plays an essential role in the above construction.

However, this is not the end of the story for the consistency of islands. The fact that island is part of the entanglement wedge of the bath subregion $R$ imposes further consistency condition. This says that operators in the island can be reconstructed as operators in the nongravitational bath. Hence a basic consistency condition is that operators in the island should be nontrivially evolved by the bath Hamiltonian as operators in the bath.\footnote{Suppose that there is no ground state degeneracy of the bath. This is also pointed out by Daniel Jafferis in a discussion.} Nevertheless, it is not obvious so far that the operator Equ.~(\ref{eq:Ophys}) has a nonzero commutator with $\hat{H}_{\text{bath}}$.

\section{The Mechanism behind the Nonlocal Information Encoding of Islands}\label{sec:mechanism}
The question we raised at the end of Sec.~\ref{sec:holography} is a basic consistency condition for the holographic interpretation of entanglement islands. Nevertheless, understanding this question is important for a proper understanding of the seemingly nonlocal information encoding scheme of islands by the bath. In this section, we will uncover the mechanism behind such information encoding scheme by answering the above question in the island model we introduced in Sec.~\ref{sec:islandmodel}.

Let's first articulate the question in the context of the explicit island model we had in Sec.~\ref{sec:islandmodel}. We proved that the gravitational AdS$_{d+1}$ is in the Higgs phase where its diffeomorphism symmetry is spontaneously broken and the graviton mass is generated via the associated Higgs mechanism. The above results are proven by firstly starting with a fully covariant description of the path integral in the gravitational AdS$_{d+1}$ and integrating out the transparent matter field $\phi_{1}(x)$ which generates a St\"{u}ckelberg mass term of the graviton in the resulting effective action. For the sake of convenience, let's consider a free massive scalar field $\phi_{R}(x)$ among the matter fields $\psi(x)$ which however obeys the standard reflective boundary condition near the asymptotic boundary of the gravitational AdS$_{d+1}$. Following Sec.~\ref{sec:holography} we have an operator $\hat{\mathcal{O}}^{\text{Phys}}(x)=\hat{\phi}_{R}(x+\hat{V}(x))$ in the island that obeys the consistency condition we proposed in \cite{Geng:2021hlu}. Nevertheless, another basic consistency condition we discussed at the end of Sec.~\ref{sec:holography} requires that
\begin{equation}
    [\hat{\mathcal{O}}^{\text{Phys}}(x),\hat{H}_{\text{bath}}]=[\hat{\phi}_{R}(x+\sqrt{16\pi G_{N}}\hat{V}(x)),\hat{H}_{\text{bath}}]\neq 0\,.\label{eq:consistency}
\end{equation}
In the dual description, we have a CFT$_{d}$ coupled to a nongravitational $(d+1)$-dimensional bath as described by Equ.~(\ref{eq:coupledH}). We studied this system perturbatively to the leading order in the coupling constant $g$. Thus we still have the result from the standard AdS/CFT correspondence that the field $\phi_{R}(x)$ is dual to a CFT$_{d}$ operator $\hat{O}_{R}(x)$ which doesn't evolve under the bath Hamiltonian. Hence we have
\begin{equation}
    [\hat{\phi}_{R}(x),\hat{H}_{\text{bath}}]= 0\,.
\end{equation}
As a result, the consistency condition Equ.~(\ref{eq:consistency}) can only be satisfied if we can show that
\begin{equation}
    [\hat{V}^{\mu}(x),\hat{H}_{\text{bath}}]\neq 0\,.\label{eq:target}
\end{equation}

\subsection{More Holography in the Island Model}\label{sec:moreholography}

To prove Equ.~(\ref{eq:target}), let us exploit a bit more about the model in Sec.~\ref{sec:islandmodel}. The nice feature of that model is that it was treated perturbatively in the coupling constant $g$ and for our purpose we only have to focus on the leading order results. The graviton in the gravitational AdS$_{d+1}$ is dual to the stress-energy tensor $T_{\text{CFT}}^{\mu\nu}(x)$ of the dual CFT$_{d}$ and the graviton mass is dual to the anomalous dimension of this stress-energy tensor. As opposed to the standard AdS/CFT, the CFT$_{d}$ is coupled to a bath, as described by Equ.~(\ref{eq:coupledH}), which implies that
\begin{equation}
    \nabla_{\mu}T^{\mu\nu}_{\text{CFT}}(x)=g \mathcal{O}_{2}(x)\partial^{\nu}\mathcal{O}_{1}(x)\,\neq0\,.\label{eq:TCFT}
\end{equation}
Thus the anomalous dimension of the CFT$_{d}$ stress-energy tensor is not protected from quantum corrections and this nonzero correction is dual to the graviton mass \cite{Aharony:2006hz}. More interestingly, Equ.~(\ref{eq:TCFT}) tells us that the CFT$_{d}$ stress-energy tensor in fact lives in a long-multiplet of the conformal group $SO(d,2)$, as opposed to conserved currents which live in short-multiplets, as it has an extra descendant which is the vector field $\mathcal{O}_{2}(x)\partial^{\nu}\mathcal{O}_{1}(x)$. For conserved currents this extra descendant corresponds to a null state by the state-operator correspondence. The dual description of this observation in the gravitational AdS$_{d+1}$ is that the graviton now lives in a long-multiplet of the AdS$_{d+1}$ isometry group $SO(d,2)$ which contains a vector field $V^{\mu}(x)$ as opposed to the short-multiplet of graviton in the standard AdS/CFT. This vector field $V^{\mu}(x)$ is exactly the St\"{u}ckelberg/Goldstone vector field we found in Equ.~(\ref{sec:islandmodel}) and this transition for graviton from being in a short-multiplet to be in a long-multiplet is exactly the Higgs mechanism that the graviton eats the vector Goldstone field $V^{\mu}(x)$ and becomes massive. The Goldstone vector field provides the extra polarization of the massive graviton as compared to the massless graviton. Therefore, we can see that the St\"{u}ckelberg/Goldstone vector field $V^{\mu}(x)$ is dual to the operator $\mathcal{O}_{2}(x)\partial^{\nu}\mathcal{O}_{1}(x)$. Before we apply this observation to the question we asked, let's work out this duality in detail. Let's define the operator $U^{\nu}(x)$ as
\begin{equation}
    \mathcal{O}_{2}(x)\partial^{\nu}\mathcal{O}_{1}(x)=\langle  \mathcal{O}_{2}(x)\partial^{\nu}\mathcal{O}_{1}(x)\rangle-U^{\nu}(x)\,,\label{eq:Udef}
\end{equation}
and we will show that $U^{\nu}(x)$ necessarily transforms non-linearly under both the bath translation and the CFT$_{d}$ translation. Under the CFT$_{d}$ translation we have
\begin{equation}
     \mathcal{O}_{2}(x)\partial^{\nu}\mathcal{O}_{1}(x)\rightarrow  \mathcal{O}_{2}(x)\partial^{\nu}\mathcal{O}_{1}(x)+ \epsilon_{\text{CFT}}^{\mu}\mathcal{O}_{2}(x)\partial^{\nu}\partial_{\mu}\mathcal{O}_{1}(x)\,.
\end{equation}
Thus, we can see that $U^{\nu}(x)$ necessarily transforms non-linearly under the CFT$_{d}$ translation and the non-linear part of the transformation is given by
\begin{equation}
    U^{\nu}(x)\rightarrow U^{\nu}(x)-\epsilon_{\text{CFT}}^{\mu}\langle\mathcal{O}_{2}(x)\partial^{\nu}\partial_{\mu}\mathcal{O}_{1}(x)\rangle\,.\label{eq:nonlinear}
\end{equation}
Before we keep going, let us explain a bit more about how we get Equ.~(\ref{eq:nonlinear}). Firstly, the condensation $\langle O_2 (x)\partial^{\nu}O_{1}(x)\rangle$ is independent on $x$ to leading nontrivial order in the coupling $g$.\footnote{To zeroth order, it is zero and to the linear order in $g$ one can confirm this using conformal perturbation theory similarly to Equ.~(\ref{eq:exporder}).} Thus, the only transformation comes from the operator piece $O_{2}(x)\partial^{\nu}O_{1}(x)$ as
\begin{equation}
\begin{split}
    U^{\nu}(x)&\rightarrow U^{\nu}(x)-\epsilon_{\text{CFT}}^{\mu}\mathcal{O}_{2}(x)\partial^{\nu}\partial_{\mu}\mathcal{O}_{1}(x)\,,\\&= U^{\nu}(x)-\epsilon_{\text{CFT}}^{\mu}\langle\mathcal{O}_{2}(x)\partial^{\nu}\partial_{\mu}\mathcal{O}_{1}(x)\rangle-\epsilon_{\text{CFT}}^{\mu}\Big(\mathcal{O}_{2}(x)\partial^{\nu}\partial_{\mu}\mathcal{O}_{1}(x)-\langle\mathcal{O}_{2}(x)\partial^{\nu}\partial_{\mu}\mathcal{O}_{1}(x)\rangle\Big)\,,\\&= U^{\nu}(x)-\epsilon_{\text{CFT}}^{\mu}\langle\mathcal{O}_{2}(x)\partial^{\nu}\partial_{\mu}\mathcal{O}_{1}(x)\rangle+\epsilon_{\text{CFT}}^{\mu}\partial_{\mu}U^{\nu}(x)\,.
    \end{split}
\end{equation}
We can ignore the $\epsilon_{\text{CFT}}^{\mu}\partial_{\mu}U^{\nu}(x)$ compared to the second term $\langle\mathcal{O}_{2}(x)\partial^{\nu}\partial_{\mu}\mathcal{O}_{1}(x)\rangle$ because it is a divergent condensation (see Equ.~(\ref{eq:exporder})).\footnote{As a trivial fact, we ignored $\epsilon_{\text{CFT}}^{\mu}\partial_{\mu}U^{\nu}(x)$ compared to $U^{\nu}(x)$ as $\epsilon^{\mu}_{\text{CFT}}(x)$ is very small.} As a result, we have Equ.~(\ref{eq:nonlinear}). The condensation can be computed to the first order in perturbation theory of the coupling $g$ as
\begin{equation}
\begin{split}
    \langle\mathcal{O}_{2}(x)\partial^{\nu}\partial^{\mu}\mathcal{O}_{1}(x)\rangle&=-\eta^{\mu\nu}\frac{ig}{d}\int d^{d}y\langle \mathcal{O}_{1}(y)\mathcal{O}_{2}(y)\mathcal{O}_{2}(x)\partial^{\rho}\partial_{\rho}\mathcal{O}_{1}(x)\rangle_{\text{CFT}}\,,\\&=-\eta^{\mu\nu}\frac{ig}{d}\int d^{d}y\frac{1}{|x-y|^{2\Delta_{2}}}\Box\frac{1}{|x-y|^{2\Delta_{1}}}\,,\\&=-\eta^{\mu\nu}\frac{ig}{d}2\Delta_{1}(2\Delta_{1}+2-d)\int d^{d}y\frac{1}{(y^2)^{1+d}}\,,\label{eq:exporder}
    \end{split}
\end{equation}
where in the last step we used the translation invariance and the fact that the double-trace deformation is marginal. The result Equ.~(\ref{eq:exporder}) is UV divergent due to the lack of a scale. Let's take a UV cutoff length scale $\epsilon$ and we parameterize Equ.~(\ref{eq:exporder}) as
\begin{equation}
      \langle\mathcal{O}_{2}(x)\partial^{\nu}\partial_{\mu}\mathcal{O}_{1}(x)\rangle=\eta^{\nu}_{\mu}\frac{1}{\epsilon^{d+2}}\,.
\end{equation}
Hence, we have the non-linear part of the transformation of $U^{\nu}(x)$ as
\begin{equation}
    U^{\nu}(x)\rightarrow U^{\nu}(x)-\epsilon^{\nu}_{\text{CFT}}\frac{1}{\epsilon^{d+2}}\,.\label{eq:UCFT}
\end{equation}
Similarly, the non-linear part of $U^{\nu}(x)$ transform under the bath translation is given by
\begin{equation}
       U^{\nu}(x)\rightarrow U^{\nu}(x)-\epsilon_{\text{bath}}^{\mu}\langle\partial_{\mu}\mathcal{O}_{2}(x)\partial^{\nu}\mathcal{O}_{1}(x)\rangle\,.
\end{equation}
To the first order in the perturbation theory of the coupling $g$, we now have
\begin{equation}
    \begin{split}
\langle\partial_{\mu}\mathcal{O}_{2}(x)\partial^{\nu}\mathcal{O}_{1}(x)\rangle&=-ig\int d^{d}y\langle\partial_{\mu}\mathcal{O}_{2}(x)\partial^{\nu}\mathcal{O}_{1}(x)\mathcal{O}_{1}(y)\mathcal{O}_{2}(y)\rangle_{\text{CFT}}\,,\\&=-\frac{4ig}{d}\eta^{\nu}_{\mu}\Delta_{1}(d-\Delta_{1})\int d^{d}y\frac{1}{(y^2)^{d+1}}\,,
    \end{split}
\end{equation}
which following our parametrization of the UV divergence becomes
\begin{equation}
    \langle\partial_{\mu}\mathcal{O}_{2}(x)\partial^{\nu}\mathcal{O}_{1}(x)\rangle=\eta^{\nu}_{\mu}\frac{2\Delta_{1}+2-d}{2(d-\Delta_{1})}\frac{1}{\epsilon^{d+2}}\,.
\end{equation}
Thus under the bath translation the non-linear part of the $U^{\nu}(x)$ transform is
\begin{equation}
    U^{\nu}(x)\rightarrow U^{\nu}(x)-\epsilon^{\nu}_{\text{bath}}\frac{2\Delta_{1}+2-d}{2(d-\Delta_{1})}\frac{1}{\epsilon^{d+2}}\,.\label{eq:Ubath}
\end{equation}
As a result, we have an operator $U^{\nu}(x)$ that transforms non-linearly under both the CFT$_{d}$ and the bath translations. Moreover, from Equ.~(\ref{eq:Udef}) we can see that the operator $U^{\nu}(x)$ has conformal weight $\Delta_{U}=d+3$. Thus, the operator $U^{\mu}(x)$ is the CFT$_{d}$ dual of the bulk Goldstone vector boson which is consistent with the fall-off behavior of $V^{\mu}(x,z)$, i.e.
\begin{equation}
   \sqrt{16\pi G_{N}} zV^{\mu}(x,z)\sim z^{d+3}(U^{\mu}(x)+O(z^2))\,,\quad\text{as }z\rightarrow0\,,\label{eq:VUdual}
\end{equation}
where we used the fact that $m_{V}^{2}=2d$ and $\mu\neq z$. As a sanity check of this result, under the bulk large diffeomorphism transform we have the non-linear part of the $V^{\mu}$ transform
\begin{equation}
    V^{\mu}(x,z)\rightarrow V^{\mu}(x,z)-\epsilon_{\text{large}}^{\mu}(x,z)\,,\quad\text{where  } \epsilon_{\text{large}}^{\mu}(x,z)\sim \frac{1}{\sqrt{16\pi G_{N}}}\Big(1+O(z^{2})\Big)\text{ as $z\rightarrow0$}\,, \label{eq:largegaugeTr}
\end{equation}
under which we have
\begin{equation}
    \delta U^{\nu}(x)\sim \frac{1}{z^{d+2}}\,,\text{ where $z\rightarrow0$}\,,
\end{equation}
which is consistent with Equ.~(\ref{eq:UCFT}) and hereafter we will take the two UV cutoff length scales to be the same, i.e. $z_{\text{UV}}=\epsilon\rightarrow0$ and identifying $\sqrt{16\pi G_{N}}\epsilon_{\text{large}}$ as $\epsilon_{\text{CFT}}$. In fact, the above match can also be checked between the $x$-dependent bulk large diffeomoprhisms and the boundary diffeomoprhisms, as we only care about the leading order term in the cutoff $\epsilon$. Moreover, the boundary operator $U^{\mu}(x)$ and its bulk dual $V^{\mu}(x)$ together with the boundary stress-energy tensor and the bulk graviton obey the standard HKLL dictionary \cite{Hamilton:2005ju,Hamilton:2006az,Hamilton:2006fh,Hamilton:2007wj,Kabat:2012av}.

\subsection{Uncovering the Mechanism}

Using the above results, we can see that
\begin{equation}
    [\sqrt{16\pi G_{N}}V^{\mu}(x,z),\hat{H}_{\text{bath}}]=[\epsilon^{d+2}U^{\mu}(x),\hat{H}_{\text{bath}}]=-i\frac{2\Delta_{1}+2-d}{2(d-\Delta_{1})}\eta^{\mu}_{0}\neq 0\,,\label{eq:nonzerocomm1}
\end{equation}
and
\begin{equation}
    [\sqrt{16\pi G_{N}}V^{\mu}(x,z),\hat{H}_{\text{CFT}_{d}}]=[\epsilon^{d+2}U^{\mu}(x),\hat{H}_{\text{CFT}_{d}}]=-i\eta^{\mu}_{0}\neq 0\,,\label{eq:nonzerocomm}
\end{equation}
where we ignored terms of higher order in $\epsilon$ and $G_{N}$.\footnote{A potential question regarding Equ.~(\ref{eq:nonzerocomm1}) and its answer are discussed in Appendix.~\ref{sec:AdSCFT}.} Thus, we have
\begin{equation}
    [\hat{O}^{\text{Phys}}(x),\hat{H}_{\text{bath}}]=[\hat{\phi}_{R}(x+\sqrt{16\pi G_{N}}\hat{V}(x)),\hat{H}_{\text{bath}}]=-i\frac{2\Delta_{1}+2-d}{2(d-\Delta_{1})}\partial_{t}\hat{O}^{\text{Phys}}(x)\neq0\,,\label{eq:bathresult}
\end{equation}
and 
\begin{equation}
    [\hat{O}^{\text{Phys}}(x),\hat{H}_{\text{CFT}_{d}}]=[\hat{\phi}_{R}(x+\sqrt{16\pi G_{N}}\hat{V}(x)),\hat{H}_{\text{CFT}_{d}}]=i\partial_{t}\hat{O}^{\text{Phys}}(x)-i\partial_{t}\hat{O}^{\text{Phys}}(x)=0\,.\label{eq:CFTresult}
\end{equation}
As a result, we proved Equ.~(\ref{eq:target}) and Equ.~(\ref{eq:consistency}). Interestingly, as a bonus we also have Equ.~(\ref{eq:CFTresult}) which is another consistency condition that operators in the island have to satisfy.\footnote{We note that $\hat{H}_{\text{CFT}_{d}}$ by itself is in fact not an observable as the coupling Equ.~(\ref{eq:coupledH}) will induce a divergent variance for the operator $\hat{H}_{\text{CFT}_{d}}$. This divergence is linear in the CFT$_{d}$ volume and so it cannot be cured by smearing. Thus counter terms have to be added to $\hat{H}_{\text{CFT}_{d}}$ such that the variance is finite and these counter terms wouldn't affect the commutator between $\hat{H}_{\text{CFT}_{d}}$ and operators in the AdS bulk on the same Cauchy slice due to causality. In fact, the ADM Hamiltonian is holographically dual to this regularized operator. Moreover, $\hat{H}_{\text{bath}}$ does not suffer from this issue, as the same divergence is localized on the boundary of the bath and we can smear the bath energy density such that its support smoothly goes to zero near the boundary.}

In fact there is a more direct way to prove Equ.~(\ref{eq:target}) which doesn't involve any explicit use of the AdS/CFT correspondence which is conceptually more illuminating but less explicit. As we have discussed in Sec.~\ref{sec:higgs}, the diffeomorphism symmetry in the gravitational AdS$_{d+1}$ part of the island model should be spontaneously broken with $V^{\mu}(x)$ as the associated Goldstone boson. It is obvious that $V^{\mu}(x)$ is a composite operator as there is no $V^{\mu}(x)$ field in the description that we start with in Sec.~\ref{sec:islandmodel}. Thus to identify this Goldstone field, we have to firstly identify the order parameter for the spontaneous diffeomorphism breaking and then recognize the Goldstone field as the diffeomorphism transform of the order parameter. In fact in our case there are infinitely many operators that serve equally good as order parameters of the spontaneous diffeomorphism breaking in the gravitational AdS$_{d+1}$. Though these operators all share a common feature that they must involve both the bath and the AdS$_{d+1}$ field and an explicit example is
\begin{equation}
    \mathcal{O}(x)=\phi_{1}(x)\phi_{2}(a)\,,\label{eq:example}
\end{equation}
where $x$ and $a$ are spacelike separated, as otherwise we cannot have position-dependent, i.e. $x$-dependent, vacuum expectation values indicating the spontaneous breaking of diffeomorphism symmetry. In the example Equ.~(\ref{eq:example}) the vacuum expectation value of the order parameter $\mathcal{O}(x)$ is given by the AdS to bath propagator \cite{Geng:2023ynk} and one can roughly identify $V^{\mu}(x)$ as
\begin{equation}
    V^{\mu}(x)\sim \phi_{2}(a)\partial^{\mu}\phi_{1}(x)\,,
\end{equation}
which is the variation of the order parameter Equ.~(\ref{eq:example}) under the diffeomorphism transform in the gravitational AdS$_{d+1}$. Thus we can see that the Goldstone field $V^{\mu}(x)$ is in fact a nonlocal operator which though acts like a local operator inside the gravitational AdS$_{d+1}$ and it acts nontrivially on the bath. As a result, we have Equ.~(\ref{eq:target}) automatically. Though we have to emphasize that this analysis only proves Equ.~(\ref{eq:target}) and it couldn't tell us what exactly $V^{\mu}(x)$ is in terms of $\phi_{1}(x)$ and $\phi_{2}(x)$. The reason is that besides Equ.~(\ref{eq:example}), there are infinitely many nonlocal operators that equally serve the role as an order parameter. An explicit construction of $V^{\mu}(x)$ to leading order in perturbation theory, which correctly transformations under the diffeomorphism, can be done following the recipe in \cite{Geng:2024dbl} component by component and the result should be consistent with the AdS/CFT consideration in the previous paragraph.

In summary, we see that the mechanism by which the information of island is encoded in the nongravitational bath, which is disconnected from the island, is that the physical operators in the island satisfying the Hamiltonian and momentum constrains and obeying the basic consistency condition in \cite{Geng:2021hlu} are in fact dressed to the nongravitational bath. This is a genuinely gravitational effect and heavily relies on the fact that graviton in the island model is massive. In the next section, we will study this mechanism in detail in a holographic island model-- the Karch-Randall braneworld \cite{Karch:2000ct,Karch:2000gx} in which we will see that the above observation that operators in the island are in fact dressed to the bath is nicely geometrized.

\section{A Holographic Toy Model-- the Karch-Randall Braneworld}\label{sec:toymodel}
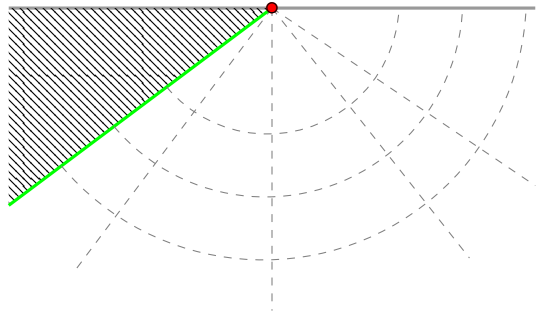
\begin{figure}
\begin{centering}
\begin{tikzpicture}[scale=1.4]
\draw[-,very thick,black!40] (-2.5,0) to (0,0);
\draw[-,very thick,black!40] (0,0) to (2.5,0);
\draw[pattern=north west lines,pattern color=black!200,draw=none] (0,0) to (-2.5,-1.875) to (-2.5,0) to (0,0);
\draw[-,dashed,color=black!50] (0,0) to (-1.875,-2.5); 
\draw[-,dashed,color=black!50] (0,0) to (0,-2.875);
\draw[-,dashed,color=black!50] (0,0) to (1.875,-2.375); 
\draw[-,dashed,color=black!50] (0,0) to (2.5,-1.6875); 
\draw[-,very thick,color=green!!50] (0,0) to (-2.5,-1.875);
\node at (0,0) {\textcolor{red}{$\bullet$}};
\node at (0,0) {\textcolor{black}{$\circ$}};
\draw[-,dashed,color=black!50] (-2,-1.5) arc (-140:-2:2.5);
\draw[-,dashed,color=black!50] (-1.5,-1.125) arc (-140:-2:1.875);
\draw[-,dashed,color=black!50] (-1,-0.75) arc (-140:-2:1.25);
\end{tikzpicture}
\caption{A constant time slice of an AdS$_{d+1}$ with a Karch-Randall brane. The green surface denotes the brane and it has AdS$_{d}$ geometry. All the dashed straight lines are in fact legitimate places for the brane to reside with the specific slice selected according to the brane tension \cite{Geng:2023qwm}. The grey-shaded region behind the brane is cutoff. The red dot is the asymptotic boundary of the brane which is also called the \textit{defect} in the literature. The defect lives at the end of the leftover asymptotic boundary (the thick black line) of the bulk.}
\label{pic:branedemo}
\end{centering}
\end{figure}

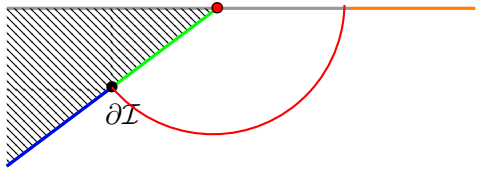
\begin{figure}
\begin{centering}
\begin{tikzpicture}[scale=1.4]
\draw[-,very thick,black!40] (-2,0) to (0,0);
\draw[-,very thick,black!40] (0,0) to (1.25,0);
\draw[-,very thick,orange] (1.25,0) to (2.45,0);
\draw[pattern=north west lines,pattern color=black!200,draw=none] (0,0) to (-2,-1.5) to (-2,0) to (0,0);
\draw[-,very thick,color=green!!50] (0,0) to (-2,-1.5);
\node at (0,0) {\textcolor{red}{$\bullet$}};
\node at (0,0) {\textcolor{black}{$\circ$}};
\draw[-,very thick,blue] (-1,-0.75) to (-2,-1.5);
\node at (-1,-0.75) {\textcolor{black}{$\bullet$}};
\node at (-0.9,-1) {\textcolor{black}{$\partial\mathcal{I}$}};
\draw[-,thick,color=red] (-1,-0.75) arc (-140:-1:1.25);
\end{tikzpicture}
\caption{A demonstration of the construction of entanglement island in the Karch-Randall braneworld. The orange line denotes the bath subregion $R$ and the blue region on the brane denotes the island $\mathcal{I}$. The red surface $\gamma$ connecting $\partial R$ and $\partial\mathcal{I}$ is a minimal area surface called the Ryu-Takayanagi surface. $\gamma$ is determined by minimizing its area also with respect to its ending point $\partial \mathcal{I}$ on the brane which correspondence to the $\min_{ \mathcal{I}}$ in the island formula Equ.~(\ref{eq:islandformula}).}
\label{pic: branwisland}
\end{centering}
\end{figure}

In this section, we will study the mechanism we uncovered in Sec.~\ref{sec:mechanism} in a strongly coupled island model. We will take the island model to be $d$-dimensional. In this model the transparent matter field is a strongly coupled CFT$_{d}$ which has a higher dimensional holographic dual. This higher dimensional holographic dual is a gravitational AdS$_{d+1}$ and it contains an end-of-the-world brane. The geometry of the end-of-the-world brane is AdS$_{d}$ which models the gravitational AdS$_{d}$ in the $d$-dimensional island model. This brane cuts off part of the bulk asymptotic boundary and intersects with the leftover asymptotic boundary. One can apply the AdS/CFT correspondence to dualize the above system to a gravitational AdS$_{d}$ coupled with a nongravitational bath where the nongravitational bath is in fact the leftover asymptotic boundary of the AdS$_{d+1}$ (see Fig.\ref{pic:branedemo}). In other words, we have an island model with a holographic dual. This system is called the Karch-Randall braneworld \cite{Karch:2000ct,Karch:2000gx,Geng:2023qwm} which plays an important role in the study of islands as islands can be easily constructed on the brane using holographic tools (see Fig.\ref{pic: branwisland}) \cite{Geng:2020qvw,Geng:2023qwm,Geng:2024xpj}.

In the Karch-Randall braneworld, the gravitational theory on the AdS$_{d}$ brane is induced from the gravitational theory in the AdS$_{d+1}$ bulk and is known to be a massive gravity theory \cite{Karch:2000ct,Karch:2000gx}. In this section, we will work out the induced gravitational theory on the brane in a fully covariant manner and we will see that the St\"{u}ckelberg vector fields on the brane are in fact gravitational Wilson lines in the AdS$_{d+1}$ bulk which connects the brane to the leftover asymptotic boundary i.e. the bath. Therefore, this is an explicit realization of the mechanism we uncovered in Sec.~\ref{sec:mechanism} that the physical operators in the island are in fact dressed to the bath. Interestingly, this mechanism is nicely geometrized in such a holographic model.

\subsection{Warm-up Exercise: Maxwell Theory}
In this subsection, we set up our notations by studying a simpler but illuminating case-- Maxwell theory living in an AdS$_{d+1}$ with a Karch-Randall brane. We will see that this simpler case inherits many properties in the gravity case but we will articulate at the end that their relevant physics in the context of islands are in fact very different.

For the sake of generality, let's consider the (d+1)-dimensional warped geometry
\begin{equation}
ds^2=d\rho^{2}+e^{2A(\rho)}\bar{g}_{ij}(x)dx^{i}dx^{j}\,,\label{eq:metric}
\end{equation}
where for AdS$_{d+1}$ we have $e^{A(\rho)}=\cosh\rho$, $\bar{g}_{ij}(x)$ is the metric for AdS$_{d}$, the brane resides along a slice $\rho=\rho_{B}$ with $\rho_{B}$ determined by the brane tension $T$ as $T=(d-1)\tanh\rho_{B}$ \cite{Geng:2023qwm} and therefore $\rho\in[\rho_{B},\infty)$. For the geometry Equ.~(\ref{eq:metric}), the nonzero components of the Chirstoffel symbol are
\begin{equation}
    \Gamma^{\rho}_{ij}=-e^{2A(\rho)}A'(\rho)\bar{g}_{ij}\,,\quad\Gamma^{i}_{\rho j}=A'(\rho)\delta^{i}_{j}\,,\quad \Gamma^{i}_{jk}=\bar{\Gamma}^{i}_{jk}\,,
\end{equation}
where $\bar{\Gamma}^{i}_{jk}$ are the Christoffel symbol for the d-dimensional metric $\bar{g}_{ij}$. Let's now study the equation of motion of the Maxwell field in this geometry. We can firstly look at the $j$-components of the (d+1)-dimensional Maxwell equation and decompose them as following
\begin{equation}
\begin{split}
\nabla_{\mu}\mathbf{F}^{\mu}_{j}=e^{-2A(r)}\bar{\nabla}_{i}\mathbf{\bar{F}}^{i}_{j}+\Big[\partial^{2}_{\rho}+(d-2)A'(\rho)\partial_{\rho}\Big]\mathbf{A}_{j}-\Big[\partial_{j}\partial_{\rho}+(d-2)A'(\rho)\partial_{j}\Big]\mathbf{A}_{\rho}\,=0,
\end{split}
\end{equation}
where $\bar{\mathbf{F}}^{i}_{j}=\bar{g}^{ik}\mathbf{F}_{kj}$. We can reorganize this equation as
\begin{equation}
\bar{\nabla}_{i}\mathbf{F}^{i}_{j}=-e^{2A(\rho)}\Big[\partial^{2}_{\rho}+(d-2)A'(\rho)\partial_{\rho}\Big]\mathbf{A}_{j}+e^{2A(\rho)}\Big[\partial_{\rho}+(d-2)A'(\rho)\Big]\partial_{j}\mathbf{A}_{\rho}\,.\label{eq:dreduce}
\end{equation}
The eigenvalues of the differential operator $e^{2A(\rho)}\Big[\partial^{2}_{\rho}+(d-2)A'(\rho)\partial_{\rho}\Big]$ are therefore the mass squares of the KK photons. Besides the equation of motion, the Maxwell field $\mathbf{A}_{\mu}$ also has to satisfy the following boundary condition close to the brane $\rho=\rho_{B}$
\begin{equation}
    \partial_{\rho}\mathbf{A}_{j}(x,\rho)|_{\rho=\rho_{B}}=0\,,\quad \mathbf{A}_{\rho}(x,\rho_{B})=0\,,\label{eq:maxwellbc}
\end{equation}
which ensures that the AdS$_{d}$ components of the Maxwell field $\mathbf{A}_{j}(x,\rho_{B})$ are free to fluctuate near the brane. With the boundary condition Equ.~(\ref{eq:maxwellbc}) imposed, we can decompose the AdS$_{d}$ components of the Maxwell field as
\begin{equation}
    \mathbf{A}_{j}(x,\rho)=\sum_{n}\psi_{n}(\rho)\bar{\mathbf{A}}^{(n)}_{j}(x)\,,
\end{equation}
where $\bar{\mathbf{A}}^{(n)}_{j}(x)$ is the $n$-th KK mode which can be thought of as a field that lives on the brane or the AdS$_{d}$ in the dual island model and $\psi_{n}(\rho)$ is the wavefunction of the $n$-th KK mode that satisfies
\begin{equation}
    -e^{2A(\rho)}\Big[\partial^{2}_{\rho}+(d-2)A'(\rho)\partial_{\rho}\Big]\psi_{n}(\rho)=m_{n}^{2}\psi_{n}(\rho)\,,\quad\partial_{\rho}\psi_{n}(\rho)|_{\rho=\rho_{B}}=0\,,\label{eq:KKwave}
\end{equation}
with $m_{n}^{2}$ denoting the square of the mass of the $n$-th KK mode and the normalization condition is
\begin{equation}
    \int_{\rho_{B}}^{\infty} d\rho e^{(d-4)A(\rho)}\psi_{n}(\rho)\psi_{m}(\rho)=\delta_{mn}\,.\label{eq:normalization}
\end{equation}
Therefore, the massless mode $\psi_{0}(\rho)=\text{const}.$ is not normalizable for $d\geq 4$ for which this massless mode would have a divergent kinetic term and thus is not physical. Furthermore, let's define
\begin{equation}
    \mathbf{\theta}(x,\rho)=-\int_{\mathcal{C}} \mathbf{A}\,,\label{eq:wilson}
\end{equation}
where $\mathcal{C}$ is an oriented contour that goes from $(x,\rho)$ to the asymptotic boundary along the $\rho$-direction. This ensures that we have
\begin{equation}
    \mathbf{A}_{\rho}(x,\rho)=\partial_{\rho}\mathbf{\theta}(x,\rho)\,.
\end{equation}
Hence we can write Equ.~(\ref{eq:dreduce}) as
\begin{equation}
\bar{\nabla}_{i}\bar{\mathbf{F}}^{i}_{j}=-e^{2A(\rho)}\Big[\partial^{2}_{\rho}+(d-2)A'(\rho)\partial_{\rho}\Big](\mathbf{A}_{j}-\partial_{j}\mathbf{\theta})\,.\label{eq:dreducetheta}
\end{equation}
As a result, we can see that if we project Equ.~(\ref{eq:dreducetheta}) to the normalizable eigenmodes $\psi_{n}(\rho)$ of the differential operator $-e^{2A(\rho)}\Big[\partial^{2}_{\rho}+(d-2)A'(\rho)\partial_{\rho}\Big]$, we get
\begin{equation}
    \bar{\nabla}_{i}\bar{\mathbf{F}}^{(n)i}_{j}(x)=m_{n}^{2}(\bar{\mathbf{A}}^{(n)}_{j}-\partial_{j}\mathbf{\theta}^{(n)})(x)\,.\label{eq:dreducethetan}
\end{equation}
The right hand side in the above equation is exactly the St\"{u}ckelberg mass term for a d-dimensional photon. Hence we expect that the dimensionally reduced theory, i.e. the gauge theory on the brane, is a tower of decoupled Maxwell theories with St\"{u}ckelberg mass terms one for each KK mode $\bar{\mathbf{A}}^{(n)}_{j}$. As a check, we can try to see if the equations of motion of the St\"{u}ckelberg fields $\theta^{(n)}(x)$ are satisfied. This can be obtained by considering the $\rho$-th component of the bulk Maxwell equation
\begin{equation}
    \begin{split}
\nabla_{\mu}\mathbf{F}^{\mu}_{\rho}&=e^{-2A(\rho)}\bar{g}^{ij}\Big[\bar{\nabla}_{i}\partial_{j}\mathbf{A}_{\rho}-\partial_{\rho}\bar{\nabla}_{j}A_{i}\Big]\,,\\&=e^{-2A(\rho)}\partial_{\rho}\Big[\bar{\nabla}^{i}\partial_{i}\mathbf{\theta}-\bar{\nabla}_{i}\mathbf{A}^{i}\Big]=0\,.\label{eq:rhomaxwell}
    \end{split}
\end{equation}
Multiplying Equ.~(\ref{eq:rhomaxwell}) by $e^{-(d-4)A(\rho)}\partial_{\rho} e^{dA(\rho)}$ we get
\begin{equation}
    e^{2A(\rho)}\Big[\partial^{2}_{\rho}+(d-2)A'(\rho)\partial_{\rho}\Big]\Big[\bar{\nabla}^{i}\partial_{i}\mathbf{\theta}-\bar{\nabla}_{i}\mathbf{A}^{i}\Big]=0\,,
\end{equation}
which gives us exactly the correct equations of motion for the tower of St\"{u}ckelberg boson fields $\theta^{(n)}(x)$ if we decompose the bulk fields $\mathbf{A}_{i}(x,\rho)$ and $\mathbf{\theta}(x,\rho)$ into the eigenmodes of the differential operator $-e^{2A(\rho)}\Big[\partial^{2}_{\rho}+(d-2)A'(\rho)\partial_{\rho}\Big]$, i.e.
\begin{equation}
m_{n}^{2}\Big[\bar{\nabla}^{i}\partial_{i}\mathbf{\theta}^{(n)}-\bar{\nabla}_{i}\mathbf{A}^{(n)i}\Big]=0\,.\label{eq:thetaeom}
\end{equation}
To obtain Equ.~(\ref{eq:thetaeom}) we have decomposed the field $\theta(x,\rho)$ as
\begin{equation}
    \theta(x,\rho)=\sum_{n}\psi_{n}(\rho)\theta^{(n)}(x)\,,
\end{equation}
for all the normalizable wavefunctions $\psi_{n}(\rho)$. We note that this decomposition is consistent with the boundary condition Equ.~(\ref{eq:maxwellbc}) i.e. 
\begin{equation}
    \partial_{\rho}\theta(x,\rho)|_{\rho_{B}}=\mathbf{A}_{\rho}(x,\rho_{B})=0\,.
\end{equation}
Moreover, the bulk gauge transform
\begin{equation}
    \mathbf{A}_{\mu}(x,\rho)\rightarrow\mathbf{A}_{\mu}(x,\rho)+\partial_{\mu}\epsilon(x,\rho)\,,\label{eq:gaugetransform}
\end{equation}
is decomposed to the gauge transforms
\begin{equation}
    \bar{\mathbf{A}}^{(n)}_{i}(x,\rho)\rightarrow\bar{\mathbf{A}}^{(n)}_{i}(x,\rho)+\partial_{i}\epsilon^{(n)}(x,\rho)\,,\quad\theta^{(n)}(x)\rightarrow\theta^{(n)}(x)+\epsilon^{(n)}(x)\,,
\end{equation}
where we have decomposed the gauge transformation parameter $\epsilon(x,\rho)$ as
\begin{equation}
    \epsilon(x,\rho)=\sum_{n}\psi_{n}(\rho)\epsilon^{(n)}(x)\,,\label{eq:epsilondecom}
\end{equation}
for all the normalizable wavefunctions $\psi_{n}(\rho)$. As a consistency check, the decomposition Equ.~(\ref{eq:epsilondecom}) for the gauge transformation parameter $\epsilon(x,\rho)$ is consistent with the fact that the boundary conditions Equ.~(\ref{eq:maxwellbc}) are preserved under the gauge transform Equ.~(\ref{eq:gaugetransform}) i.e.
\begin{equation}
    \partial_{\rho}\epsilon(x,\rho)|_{\rho=\rho_{B}}=0\,,
\end{equation}
also with $\epsilon(x,\rho)\rightarrow0$ as $\rho\rightarrow\infty$.

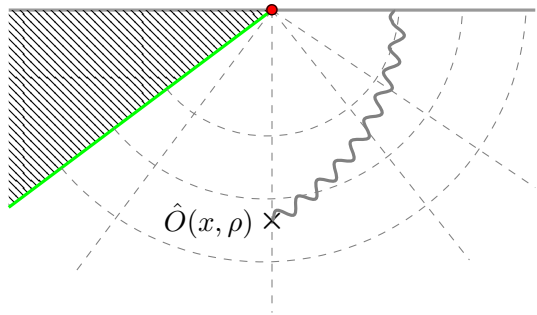
\begin{figure}
\begin{centering}
\begin{tikzpicture}[scale=1.4]
\draw[-,very thick,black!40] (-2.5,0) to (0,0);
\draw[-,very thick,black!40] (0,0) to (2.5,0);
\draw[pattern=north west lines,pattern color=black!200,draw=none] (0,0) to (-2.5,-1.875) to (-2.5,0) to (0,0);
\draw[-,dashed,color=black!50] (0,0) to (-1.875,-2.5); 
\draw[-,dashed,color=black!50] (0,0) to (0,-2.875);
\draw[-,dashed,color=black!50] (0,0) to (1.875,-2.375); 
\draw[-,dashed,color=black!50] (0,0) to (2.5,-1.6875); 
\draw[-,very thick,color=green!!50] (0,0) to (-2.5,-1.875);
\node at (0,0) {\textcolor{red}{$\bullet$}};
\node at (0,0) {\textcolor{black}{$\circ$}};
\draw[-,dashed,color=black!50] (-2,-1.5) arc (-140:-2:2.5);
\draw[-,dashed,color=black!50] (-1.5,-1.125) arc (-140:-2:1.875);
\draw[-,dashed,color=black!50] (-1,-0.75) arc (-140:-2:1.25);
\node at (0,-2) {\textcolor{black}{$\cross$}};
\node at (-0.6,-2) {\textcolor{black}{$\hat{O}(x,\rho)$}};
\draw[-,very thick,snake it,color=black!50] (0,-2) arc (-90:0:2);
\end{tikzpicture}
\caption{A demonstration of the dressing using the Wilson line $\theta(x,\rho)$. The cross denotes the operator insertion at $(x,\rho)$ and the wavy curve connecting the cross to the asymptotic boundary denotes the Wilson line. This Wilson line can end at any point on the asymptotic boundary}
\label{pic:bulkthetadress}
\end{centering}
\end{figure}

\begin{figure}
\begin{centering}
\begin{tikzpicture}[scale=1.4]
\draw[-,very thick,black!40] (-2.5,0) to (0,0);
\draw[-,very thick,black!40] (0,0) to (2.5,0);
\draw[pattern=north west lines,pattern color=black!200,draw=none] (0,0) to (-2.5,-1.875) to (-2.5,0) to (0,0);
\draw[-,dashed,color=black!50] (0,0) to (-1.875,-2.5); 
\draw[-,dashed,color=black!50] (0,0) to (0,-2.875);
\draw[-,dashed,color=black!50] (0,0) to (1.875,-2.375); 
\draw[-,dashed,color=black!50] (0,0) to (2.5,-1.6875); 
\draw[-,very thick,color=green!!50] (0,0) to (-2.5,-1.875);
\node at (0,0) {\textcolor{red}{$\bullet$}};
\node at (0,0) {\textcolor{black}{$\circ$}};
\draw[-,dashed,color=black!50] (-2,-1.5) arc (-140:-2:2.5);
\draw[-,dashed,color=black!50] (-1.5,-1.125) arc (-140:-2:1.875);
\draw[-,dashed,color=black!50] (-1,-0.75) arc (-140:-2:1.25);
\node at (-1.25,-0.9375) {\textcolor{black}{$\cross$}};
\node at (-1.15,-1.3) {\textcolor{black}{$\hat{O}(x)$}};
\draw[-,very thick,snake it,color=black!50] (-1.25,-0.9375) arc (-135:-5:1.56);
\end{tikzpicture}
\caption{A demonstration of the dressing using the St\"uckelberg fields $\theta^{(n)}(x)$ for charged operators on the brane. The St\"{u}ckelberg fields $\theta^{(n)}(x)$ are the KK modes of the bulk Wilson line $\theta(x,\rho)$ i.e. $\theta^{(n)}(x)=\int_{\rho_{B}}^{\infty}e^{(d-4)A(\rho)}\psi_{n}(\rho)\theta(x,\rho)$. Therefore, $\theta^{(n)}(x)$ can be thought as a smeared bulk Wilson line with an end point at $(x,\rho_{B})$ on the brane and another end on the asymptotic boundary of the bulk. This smeared Wilson line goes purely in the $\rho$-direction.}
\label{pic:branethetadress}
\end{centering}
\end{figure}
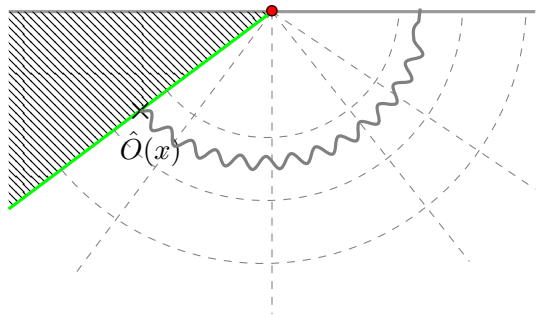

The interesting point of the above analysis is that the St\"{u}ckelberg boson field $\mathbf{\theta}$ is in fact the bulk Wilson line Equ.~(\ref{eq:wilson}) which transforms under the gauge transform Equ.~(\ref{eq:gaugetransform}) as
\begin{equation}
    \theta(x,\rho)\rightarrow\theta(x,\rho)+\epsilon(x,\rho)\,.
\end{equation}
Thus $\theta(x,\rho)$ can be used to dress any bulk charged operator to a gauge invariant operator (see Fig.\ref{pic:bulkthetadress}). For example, for a charged operator $\hat{O}(x,\rho)$ in the bulk that transforms as
\begin{equation}
    \hat{O}(x,\rho)\rightarrow e^{iq\epsilon(x,\rho)}\hat{O}(x,\rho)\,,
\end{equation}
we can get a gauge invariant operator
\begin{equation}
    \hat{O}^{\text{Phys}}(x,\rho)=e^{-iq\theta(x,\rho)}\hat{O}(x,\rho)\,.
\end{equation}
Similarly, in the induced theory on the brane, the Goldstone boson fields $\theta^{(n)}(x)$ can be used to dress any operators charged under $\bar{\mathbf{A}}^{(n)}_{i}(x)$ to a gauge invariant operator. For instance, if there exists an operator $\hat{O}(x)$ that is charged under $\bar{\mathbf{A}}^{(n)}_{i}(x)$ as
\begin{equation}
    \hat{O}(x)\rightarrow e^{iq_{(n)}\epsilon^{n}(x)}\hat{O}(x)\,,
\end{equation}
then we can construct a gauge invariant operator
\begin{equation}
    \hat{O}^{\text{Phys}}(x)=e^{-iq_{n}\theta^{(n)}(x)}\hat{O}(x)\,.\label{eq:branedresstheta}
\end{equation}
Intuitively, one can understand Equ.~(\ref{eq:branedresstheta}) as the statement that in the dual island model charged operators in the AdS$_{d}$, i.e. the brane, can be dressed to the bath, i.e. the asymptotic boundary of the bulk AdS$_{d+1}$ (see Fig.\ref{pic:branethetadress}). To put this intuition on a firm ground, we have to show that the dressed operator Equ.~(\ref{eq:branedresstheta}) can be detected on the bath, i.e. the asymptotic boundary on the bulk AdS$_{d+1}$. This can be easily shown and it is in fact just the Gauss' law. The charge operator
\begin{equation}
    \hat{Q}=-\int d^{d-1}\vec{x}\sqrt{-\bar{g}}e^{(d-2)A(\rho_{c})} \mathbf{F}_{0\rho}(t,\vec{x},\rho_{c})\,,\quad \rho_{c}\rightarrow\infty\,,
\end{equation}
is fully accessible to the asymptotic boundary of the bulk AdS$_{d+1}$ and is hence a bath operator. In the AdS$_{d+1}$ bulk we have the equal-time commutation relation
\begin{equation}
    [\mathbf{A}_{\rho}(t,\vec{x},\rho),\mathbf{F}_{0\rho}(t,\vec{y},\rho')]=i\frac{1}{\sqrt{-\bar{g}}e^{(d-2)A(\rho)}}\delta(\rho-\rho')\delta^{d-1}(\vec{x}-\vec{y})\,,
\end{equation}
from canonical quantization. Thus we have\footnote{Note that the integral $\int_{\rho_{B}}^{\infty} d\rho e^{(d-4)A(\rho)}\psi_{n}(\rho)$ is finite for normalizable modes as one can show from Equ.~(\ref{eq:KKwave}) that as $\rho\rightarrow\infty$ we have $\psi_{n}(\rho)\sim e^{-(d-2)A(\rho)}=\cosh^{-(d-2)}\rho$.}
\begin{equation}
    [\hat{Q},\theta(x,\rho)]=-i\,\Rightarrow[\hat{Q},\theta^{(n)}(x)]=-i\int_{\rho_{B}}^{\infty} d\rho e^{(d-4)A(\rho)}\psi_{n}(\rho)\neq0\,,
\end{equation} 
which implies that for Equ.~(\ref{eq:branedresstheta})
\begin{equation}
    [\hat{Q},\hat{O}^{\text{Phys}}(x)]=[\hat{Q},e^{-iq\theta^{(n)}(x)}]\hat{O}(x)\neq0,
\end{equation}
where the first step used the locality for spatially separated operators. As a result, we indeed see that the dressed operator Equ.~(\ref{eq:branedresstheta}) is detectable on the bath and the bath observer can detect it using the charge operator $\hat{Q}$. 

Nevertheless, we notice that there also exist gauge invariant operators that cannot be detected by the charge operator $\hat{Q}$. For example, the local electric field strength $\vec{E}(x,\rho)$ in the bulk or $\vec{E}^{(n)}(x)$ on the brane or operators of the type $\phi(x,\rho)W(x,\rho;y,\rho')\phi^{*}(y,\rho)$ where $W(x,\rho;y,\rho')=e^{iq\int_{\mathcal{C}_{xy}}\mathbf{A}}$ is the Wilson line associated with the path $\mathcal{C}_{xy}$ connecting the bulk points $(x,\rho)$ and $(y,\rho')$ with $\phi(x,\rho)$ an operator of charge $q$. As we will see in the next subsection, this is one of the distinctions between gauge theory and gravity as in gravity the analogous of charge is the energy and one doesn't have negatively energetic operators for a quantum system whose energy is bounded from below. Hence there don't exist operators in the AdS$_{d+1}$ bulk or on the brane, i.e. in the gravitational AdS$_{d}$ in the dual island model, that are not detectable by observers in the bath.

\subsection{Gravity}\label{sec:gravKR}
In this subsection, we consider the fully-fledged gravitational case in the Karch-Randall braneworld. For our purpose, we will firstly generalize the original analysis in \cite{Karch:2000ct} to a fully covariant form. Then we will explicitly construct operators in the entanglement island that obey the consistency condition proposed in \cite{Geng:2021hlu} and study how these operators satisfy the new consistency condition we proposed in Equ.~(\ref{eq:consistency}). We will see that the Karch-Randall braneworld provides the holographic dual of the mechanism we uncovered in Sec.~\ref{sec:mechanism}.

\subsubsection{The Theory on the Brane}
The linearized Einstein's equation in the AdS$_{d+1}$ bulk is
\begin{equation}
    \frac{1}{2}g_{\mu\nu}\nabla_{\alpha}\nabla_{\beta}h^{\alpha\beta}-\frac{1}{2}g_{\mu\nu}\Box h+\frac{1}{2}\nabla_{\mu}\nabla_{\nu}h-\frac{1}{2}\nabla_{\alpha}\nabla_{\mu}h_{\nu}^{\alpha}-\frac{1}{2}\nabla_{\alpha}\nabla_{\nu}h_{\mu}^{\alpha}+\frac{1}{2}\Box h_{\mu\nu}+\frac{d}{2}(g_{\mu\nu}h-2h_{\mu\nu})\,,\label{eq:linearizedEF}
\end{equation}
where $g_{\mu\nu}$ denotes the bulk background metric and we have used the fact that the cosmological constant is $-\frac{d(d-1)}{2}$ with the AdS length scale set to one. The fundamental field is the metric fluctuation, i.e. the graviton field, $h_{\mu\nu}$ and all upper indices are induced by contracting the low indices with the inverse background metric $g^{\mu\nu}$. 

We want to decompose Equ.~(\ref{eq:linearizedEF}) into relevant components in the background metric Equ.~(\ref{eq:metric}) with $e^{A(\rho)}=\cosh\rho$ and $\bar{g}_{ij}(x)$ as the metric for empty AdS$_{d}$. The $ij$-components of Equ.~(\ref{eq:linearizedEF}) are
\begin{equation}
    \begin{split}
        &\frac{1}{2}\bar{g}_{ij}\bar{\nabla}_{m}\bar{\nabla}_{n}\bar{h}^{mn}-\frac{1}{2}\bar{g}_{ij}\text{\ovA{$\Box$}}\bar{h}+\frac{1}{2}\bar{\nabla}_{i}\bar{\nabla}_{j}\bar{h}-\frac{1}{2}\bar{\nabla}_{k}\bar{\nabla}_{i}\bar{h}_{j}^{k}-\frac{1}{2}\bar{\nabla}_{k}\bar{\nabla}_{j}\bar{h}_{i}^{k}+\frac{1}{2} \text{\ovA{$\Box$}}h_{ij}+\frac{d-1}{2}(\bar{g}_{ij}\bar{h}-2h_{ij})\,\\&+\frac{1}{2}e^{-(d-4)A(\rho)}\partial_{\rho}\Big[e^{dA(\rho)}\partial_{\rho}e^{-2A(\rho)}(h_{ij}-\bar{g}_{ij}\bar{h})\Big]+e^{2A(\rho)}\bar{g}_{ij}(d-2)A'(\rho)\bar{\nabla}_{i}\bar{h}^{\rho i}+e^{2A(\rho)}\bar{g}_{ij}\bar{\nabla}_{j}\partial_{\rho}\bar{h}_{\rho}^{j}\,\\&-\frac{e^{2A(\rho)}}{2}\bar{\nabla}_{i}\partial_{\rho}h_{j}^{\rho}-\frac{2A(\rho)}{2}\bar{\nabla}_{j}\partial_{\rho}h_{i}^{\rho}-\frac{d-2}{2}e^{2A(\rho)}A'(\rho)\bar{\nabla}_{i}h_{j}^{\rho}-\frac{d-2}{2}e^{2A(\rho)}A'(\rho)\bar{\nabla}_{j}h_{i}^{\rho}\\&+\frac{e^{4A(\rho)}\bar{g}_{ij}}{2}\Big[(d-1)A'(\rho)\partial_{\rho}+d(d-2)A'^{2}(\rho)+(d-2)A''(\rho)-e^{-2A(\rho)}\text{\ovA{$\Box$}}+\frac{d}{2}\Big]h_{\rho\rho}+\frac{e^{2A(\rho)}}{2}\bar{\nabla}_{i}\partial_{j}h_{\rho\rho}\,,\label{eq:branegravitoneom}
    \end{split}
\end{equation}
where $\bar{\nabla}_{i}$ is the torsion free and metric compatiable derivative with respect to $\bar{g}_{ij}(x)$, the upper indices of $\bar{h}^{mn}$ are induced by the inverse metric $\bar{g}^{ij}$, $\bar{h}$ is the trace of $h_{ij}$ under the metric $\bar{g}_{ij}$ and the $\rho$-indices of $h_{\mu\nu}$ are scalar under the derivative $\bar{\nabla}_{i}$. To obtain Equ.~(\ref{eq:branegravitoneom}) from the bulk linearized Einsteins's equation Equ.~(\ref{eq:linearizedEF}), we have stripped off a common factor $e^{-2A(\rho)}$ for each term. The above equation is not in a particularly nice form due to the $h^{\rho}_{i}$ and $h_{\rho\rho}$ terms. However, one should notice that it is invariant under the gauge transform
\begin{equation}
    h_{\mu\nu}(x,\rho)\rightarrow h_{\mu\nu}(x,\rho)+\nabla_{\mu}\epsilon_{\nu}(x,\rho)+\nabla_{\nu}\epsilon_{\mu}(x,\rho)\,,\label{eq:diffeo}
\end{equation}
so we can transform Equ.~(\ref{eq:branegravitoneom}) into a nice form by appropriate field redefinitions. Let's first define the vector field $V_{\mu}(x,\rho)$ whose components are
\begin{equation}
    \begin{split}
   V_{\rho}(x,\rho)&=\frac{1}{2}\int_{\rho}^{\infty} du h_{\rho\rho}(x,u)\,, \\ V_{i}(x,\rho)&=e^{2A(\rho)}\int_{\rho}^{\infty} du e^{-2A(u)}h_{\rho i}(x,u)+\frac{e^{2A(\rho)}}{2}\int _{\rho}^{\infty} due^{-2A(u)}\int_{u}^{\infty} du' \partial_{i}h_{\rho\rho}(x,u')\,.\label{eq:Vdef}
    \end{split}
\end{equation}
This vector is designed such that we have
\begin{equation}
    h_{\rho\mu}(x,\rho)=-\nabla_{\rho}V_{\mu}(x,\rho)-\nabla_{\mu}V_{\rho}(x,\rho)\,,
\end{equation}
and this vector is in fact the gravitational Wilson line which transforms under the gauge transform Equ.~(\ref{eq:diffeo}) as
\begin{equation}
    V_{\mu}(x,\rho)\rightarrow V_{\mu}(x,\rho)-\epsilon_{\mu}(x,\rho)\,.\label{eq:Vtrans}
\end{equation}
Now we can use this vector field to simplify Equ.~(\ref{eq:branegravitoneom}) if we do the following field redefinition
\begin{equation}
    \tilde{h}_{ij}(x,\rho)= h_{ij}(x,\rho)+2e^{2A(\rho)}A'(\rho)\bar{g}_{ij}(x)V_{\rho}(x,\rho)\,.\label{eq:newh}
\end{equation}
This new field is invariant under the transform Equ.~(\ref{eq:diffeo}) generated by the $\rho$-direction deformation $\epsilon_{\rho}(x,\rho)$ and it transforms under the $i$-direction deformations as
\begin{equation}
    \tilde{h}_{ij}(x,\rho)\rightarrow \tilde{h}_{ij}(x,\rho)+\bar{\nabla}_{i}\epsilon_{j}(x,\rho)+\bar{\nabla}_{j}\epsilon_{i}(x,\rho)\,,
\end{equation}
which makes it a better candidate for a $d$-dimensional graviton field. Using this new field and the definition Equ.~(\ref{eq:Vdef}) we can simplify Equ.~(\ref{eq:branegravitoneom}) as 
\begin{equation}
    \begin{split}
        0=&\frac{1}{2}\bar{g}_{ij}\bar{\nabla}_{m}\bar{\nabla}_{n}\tilde{h}^{mn}-\frac{1}{2}\bar{g}_{ij}\text{\ovA{$\Box$}} \tilde{h}+\frac{1}{2}\bar{\nabla}_{i}\bar{\nabla}_{j}\tilde{h}-\frac{1}{2}\bar{\nabla}_{k}\bar{\nabla}_{i}\tilde{h}_{j}^{k}-\frac{1}{2}\bar{\nabla}_{k}\bar{\nabla}_{j}\tilde{h}_{i}^{k}+\frac{1}{2} \text{\ovA{$\Box$}}\tilde{h}_{ij}+\frac{d-1}{2}(\bar{g}_{ij}\tilde{h}-2\tilde{h}_{ij})\,\\+\frac{1}{2}&e^{-(d-4)A(\rho)}\partial_{\rho}\Big[e^{dA(\rho)}\partial_{\rho}e^{-2A(\rho)}(\tilde{h}_{ij}-\bar{g}_{ij}\tilde{h})\Big]+\frac{1}{2}e^{-(d-4)A(\rho)}\partial_{\rho}\Big[e^{dA(\rho)}\partial_{\rho}e^{-2A(\rho)}(\bar{\nabla}_{i}V_{j}+\bar{\nabla}_{j}V_{i}-\bar{g}_{ij}2\bar{\nabla}_{i}V^{i})\Big]\,,\label{eq:branegravitoneomsimplify}
    \end{split}
\end{equation}
where the upper indices are all lifted from lower indices using $\bar{g}^{ij}$. Similarly, in terms of the new field Equ.~(\ref{eq:newh}) and the vector field $V_{i}$, we have the $i\rho$-components of the bulk linearized Einstein equation Equ.~(\ref{eq:linearizedEF})
\begin{equation}
    \frac{1}{2}\partial_{\rho}\Big[e^{-2A(\rho)}\big[\partial_{i}\tilde{h}-\bar{\nabla}^{j}\tilde{h}_{ij}-\bar{\nabla}_{j}\Big(\bar{\nabla}_{i}V^{j}+\bar{\nabla}^{j}V_{i}\Big)+2\bar{\nabla}_{i}\bar{\nabla}_{j}V^{j}\big]\Big]=0\,,
\end{equation}
with all upper indices lifted from lower ones using $\bar{g}^{ij}$. This equation, after multiplying $-e^{-(d-4)A(\rho)}\partial_{\rho}e^{dA(\rho)}$, can be equivalently written as
\begin{equation}
    \frac{1}{2}e^{-(d-4)A(\rho)}\partial_{\rho}\Big[e^{dA(\rho)}\partial_{\rho}e^{-2A(\rho)}\big[\bar{\nabla}^{j}\big(\tilde{h}_{ij}+\bar{\nabla}_{i}V_{j}+\bar{\nabla}_{j}V_{i}\big)-\bar{\nabla}_{i}\big(\tilde{h}+2\bar{\nabla}_{j}V^{j}\big)\big]\Big]=0\,.\label{eq:GoldstonVeom}
\end{equation}
Moreover, the $\rho\rho$ component of the bulk Einstein's equation Equ.~(\ref{eq:linearizedEF}) can be combined with Equ.~(\ref{eq:branegravitoneomsimplify}) to get
\begin{equation}
    \frac{1}{2}e^{-(d-4)A(\rho)}\partial_{\rho}\Big[e^{dA(\rho)}\partial_{\rho}e^{-2A(\rho)}(\tilde{h}+2\bar{\nabla}_{i}V^{i})\Big]=0\,.\label{eq:rhorho}
\end{equation}
We note that the resulting three equations Equ.~(\ref{eq:branegravitoneomsimplify}), Equ.~(\ref{eq:GoldstonVeom}) and Equ.~(\ref{eq:rhorho}) doesn't contain the vector field component $V_{\rho}(x,\rho)$. This can be understood by noticing that the bulk Einstein's equation Equ.~(\ref{eq:linearizedEF}) is manifestly invariant under the transform Equ.~(\ref{eq:diffeo}) together with the following considerations. In terms of the new field Equ.~(\ref{eq:newh}) and the vector field $V_{\mu}(x,\rho)$, the gauge transform Equ.~(\ref{eq:diffeo}) is equivalently
\begin{equation}
    \begin{split}
        \tilde{h}_{ij}(x,\rho)&\rightarrow \tilde{h}_{ij}(x,\rho)+\bar{\nabla}_{i}\epsilon_{j}(x,\rho)+\bar{\nabla}_{j}\epsilon_{i}(x,\rho)\,,\\V_{i}(x,\rho)&\rightarrow V_{i}(x,\rho)-\epsilon_{i}(x,\rho)\,,\quad V_{\rho}(x,\rho)\rightarrow V_{\rho}(x,\rho)-\epsilon_{\rho}(x,\rho)\,.\label{eq:transnew}
    \end{split}
\end{equation}
Therefore, the dimensionally reduced Einstein's equations must be linear in the above fields and invariant under the gauge transformations Equ.~(\ref{eq:transnew}) and so $V_{\rho}(x,\rho)$ must manifestly disappear from these equations as none of the other fields transform under $\epsilon_{\rho}(x,\rho)$.

The three equations Equ.~(\ref{eq:branegravitoneom}), Equ.~(\ref{eq:GoldstonVeom}) and Equ.~(\ref{eq:rhorho}) are manifestly invariant under the transformation Equ.~(\ref{eq:diffeo}), or equivalently Equ.~(\ref{eq:transnew}), and each term in these three equations in fact has a nice interpretation if we notice that the eigenvalues of the differential operator $-e^{-(d-4)A(\rho)}\partial_{\rho}e^{dA(\rho)}\partial_{\rho}e^{-2A(\rho)}$ are the mass of the KK gravitons \cite{Karch:2000ct}. Let $\phi_{n}(\rho)$ be the KK wavefunctions with a Neumann boundary condition on the brane, i.e. eigenfunctions of the differential operator $-e^{-(d-4)A(\rho)}\partial_{\rho}e^{dA(\rho)}\partial_{\rho}e^{-2A(\rho)}$, that satisfy
\begin{equation}
    -e^{-(d-4)A(\rho)}\partial_{\rho}e^{dA(\rho)}\partial_{\rho}e^{-2A(\rho)} \phi_{n}(\rho)=m_{n}^{2}\phi_{n}(\rho)\,,\quad\partial_{\rho}e^{-2A(\rho)}\phi_{n}(\rho)|_{\rho=\rho_{B}}=0\,.\label{eq:KKeomgraviton}
\end{equation}
From this equation, the orthognality condition of the wavefunctions $\phi_{n}(\rho)$ is
\begin{equation}
    \int_{\rho_{B}}^{\infty} d\rho e^{(d-6)A(\rho)}\phi_{n}(\rho)\phi_{m}(\rho)=\delta_{nm}\,,\label{eq:normgraviton}
\end{equation}
where we also normalized the wavefunctions. Thus we can see that the zero mass eigenmode $\phi_{0}(\rho)=e^{2A(\rho)}=\cosh^{2}\rho$ is not normalizable for $d\geq2$ which includes all cases with a dynamical graviton in the bulk. Hence the physical modes are all massive with nonzero mass squares $m_{n}^{2}$. As a result, if we project all the fields to the $r$-th eigenmode then we have
\begin{equation}
    \begin{split}
        &\frac{1}{2}\bar{g}_{ij}\bar{\nabla}_{m}\bar{\nabla}_{n}\tilde{h}^{(r)mn}-\frac{1}{2}\bar{g}_{ij}\text{\ovA{$\Box$}}\tilde{h}^{(r)}+\frac{1}{2}\bar{\nabla}_{i}\bar{\nabla}_{j}\tilde{h}^{(r)}-\frac{1}{2}\bar{\nabla}_{k}\bar{\nabla}_{i}\tilde{h}_{j}^{(r)k}-\frac{1}{2}\bar{\nabla}_{k}\bar{\nabla}_{j}\tilde{h}_{i}^{(r)k}+\frac{1}{2} \text{\ovA{$\Box$}}\tilde{h}_{ij}^{(r)}+\frac{d-1}{2}(\bar{g}_{ij}\tilde{h}^{(r)}-2\tilde{h}_{ij}^{(r)})\,\\&-\frac{m_{r}^{2}}{2}(\tilde{h}_{ij}^{(r)}-\bar{g}_{ij}\tilde{h}^{(r)})-\frac{m_{r}^{2}}{2}(\bar{\nabla}_{i}V_{j}^{(r)}+\bar{\nabla}_{j}V_{i}^{(r)}-\bar{g}_{ij}2\bar{\nabla}_{i}V^{(r)i})=0\,,
        \label{eq:branegravitoneomKK}
    \end{split}
\end{equation}
and
\begin{equation}
    \frac{m_{r}^{2}}{2}\Big[\bar{\nabla}^{j}\big(\tilde{h}^{(r)}_{ij}+\bar{\nabla}_{i}V^{(r)}_{j}\bar{\nabla}_{j}V^{(r)}_{i}\big)-\bar{\nabla}_{i}\big(\tilde{h}^{(r)}+2\bar{\nabla}_{j}V^{(r)j}\big)\Big]=0\,,\label{eq:VeomKK}
\end{equation}
together with
\begin{equation}
 m_{r}^{2}[\tilde{h}^{(r)}+2\bar{\nabla}_{i}V^{(r)i}]=0\,,\label{eq:rhorhoKK}   
\end{equation}
where $m_{r}^{2}$ is the mass square of the $r$-th KK mode and all the fields $\tilde{h}^{(r)}$ and $V^{(r)i}$ are only functions of $x$, i.e. they are d-dimensional fields. Therefore, one can recognize that the terms on the first row of Equ.~(\ref{eq:branegravitoneomKK}) are standard linearized Einstein's equation in AdS$_{d}$, the first term in the second line of Equ.~(\ref{eq:branegravitoneomKK}) is exactly the Fierz-Pauli mass term for graviton \cite{Hinterbichler:2011tt} and the second term with the field $V^{(r)}_{i}$ is exactly of the St\"{u}ckelberg form which ensures that the graviton mass term wouldn't manifestly break the gauge symmetry Equ.~(\ref{eq:diffeo}). Moreover, Equ.~(\ref{eq:VeomKK}) is the equation of motion from the variation of the graviton Fierz-Pauli mass term in the St\"{u}ckelberg form with respect to the Goldstone vector $V^{(r)}_{i}(x)$ field. Finally, Equ.~(\ref{eq:rhorhoKK}) is the statement that the diffeomorphism invariant combinations $\tilde{h}_{ij}^{(r)}+\bar{\nabla}_{i}V^{(r)}_{j}+2\bar{\nabla}_{j}V^{(r)}_{i}$ are traceless.

Nevertheless, for the above projection to be consistent, we also have to prove that $\tilde{h}_{ij}(x,\rho)$ and $V_{i}(x,\rho)$ obeys the same Neumann boundary condition as the wavefunctions $\phi_{n}(\rho)$ near the brane. This can be done by exploiting the gauge invariance Equ.~(\ref{eq:transnew}). As it is standard in the Karch-Randall braneworld, the geometry of the brane is determined by the brane embedding equation
\begin{equation}
    K_{\mu\nu}-g^{B}_{\mu\nu}K+Tg^{B}_{\mu\nu}=0\,,\label{eq:braneembedding}
\end{equation}
where $K_{\mu\nu}$ is the extrinsic curvature of the brane, $g^{B}_{\mu\nu}$ is the induced metric on the brane and $T$ is the tension of the brane. The boundary condition of the bulk metric fluctuation $h_{\mu\nu}$ is determined by linearizing Equ.~(\ref{eq:braneembedding}). This is a complicated equation which though is diffeomorphism invariant, i.e. invariant under Equ.~(\ref{eq:diffeo}), if the background brane induced metric satisfies Equ.~(\ref{eq:braneembedding}). Therefore, one can exploit this invariance to simplify the analysis of the boundary condition. One can firstly fix the gauge parameter $\epsilon_{\rho}(x,\rho)$ by setting
\begin{equation}
    V_{\rho}(x,\rho)=0\,.
\end{equation}
Then one can fix the other gauge parameters $\epsilon_{i}(x,\rho)$ by setting
\begin{equation}
    V_{i}(x,\rho)=0\,.
\end{equation}
These choices completely fix all the gauge. Within this gauge, it is easy to linearize Equ.~(\ref{eq:braneembedding}) which is equivalent to
\begin{equation}
    \delta K_{ij}-\tilde{h}_{ij} K+T \tilde{h}_{ij}|_{\rho=\rho_{B}}=0\,,\label{eq:bcbrane1}
\end{equation}
Using $K_{ij}=\frac{1}{2}\partial_{\rho} g_{ij}$, which is correct within our gauge, we get
\begin{equation}
    \begin{split}
        T=\frac{d-1}{d}K=(d-1)A'(\rho)\,,
        \end{split}
\end{equation}
which simplifies Equ.~(\ref{eq:bcbrane1}) to
\begin{equation}
    \frac{1}{2}\partial_{\rho}\tilde{h}_{ij}-A'(\rho)\tilde{h}_{ij}=\frac{e^{2A(\rho)}}{2}\partial_{\rho}e^{-2A(\rho)}\tilde{h}_{ij}(x,\rho)|_{\rho=\rho_{B}}=0\,.
\end{equation}
As a result, due to $e^{2A(\rho_{B})}\neq0$, the metric fluctuation near the brane exactly satisfies the same boundary condition as the wavefunctions $\phi_{n}(\rho)$\,. Restoring the gauge, we get the fully covariant boundary conditions
\begin{equation}
    \partial_{\rho}e^{-2A(\rho)}\tilde{h}_{ij}(x,\rho)_{\rho=\rho_{B}}=0\,,\quad\partial_{\rho}e^{-2A(\rho)}V_{i}(x,\rho)|_{\rho=\rho_{B}}=0\,,\quad\partial_{\rho}A'(\rho)V_{\rho}(x,\rho)|_{\rho=\rho_{B}}=0\,,\label{eq:bccovariant}
\end{equation}
and this boundary condition requires that the gauge parameter $\epsilon_{\mu}(x,\rho)$ satisfy the same boundary conditions
\begin{equation}
    \partial_{\rho}e^{-2A(\rho)}\epsilon_{i}(x,\rho)|_{\rho=\rho_{B}}=0\,,\quad\partial_{\rho}A'(\rho)\epsilon_{\rho}(x,\rho)|_{\rho=\rho_{B}}=0\,.
\end{equation}
Therefore, the projection of the fields $\tilde{h}_{ij}$ and $V_{i}$ to the KK modes in obtaining Equ.~(\ref{eq:branegravitoneomKK}) and Equ.~(\ref{eq:VeomKK}) is a consistent operation. 

In summary, at the linearized level, the gravitational theory on the brane consists of a tower of decoupled massive gravity theories with gravitons $\tilde{h}^{(r)}_{ij}(x)$ and St\"{u}ckelberg vector fields $V^{(r)}_{i}(x)$. Moreover, each massive gravity sector has its own spontaneously broken diffeomorphism symmetry
\begin{equation}
       \tilde{h}^{(r)}_{ij}(x)\rightarrow \tilde{h}^{(r)}_{ij}(x)+\bar{\nabla}_{i}\epsilon^{(r)}_{j}(x)+\bar{\nabla}_{j}\epsilon^{(r)}_{i}(x)\,,\quad V^{(r)}_{i}(x)\rightarrow V^{(r)}_{i}(x)-\epsilon^{(r)}_{i}(x)\,.
\end{equation}
The relations between the above brane fields and diffeomorphism parameter and the bulk fields and diffeomorphism parameter are
\begin{equation}
    \tilde{h}_{ij}(x,\rho)=\sum_{r}\phi_{r}(\rho)\tilde{h}^{(r)}_{ij}(x)\,,\quad V_{i}(x,\rho)=\sum_{r}\phi_{r}(\rho) V_{i}(x)\,,\quad \epsilon_{i}(x,\rho)=\sum_{r}\phi_{r}(\rho)\epsilon^{(r)}_{i}(x)\,,
\end{equation}
where $\phi_{r}(\rho)$ is the $r$-th KK wavefunction following Equ.~(\ref{eq:KKeomgraviton}), $V_{i}(x,\rho)$ are in fact the bulk gravitational Wilson lines Equ.~(\ref{eq:Vdef}) and $\tilde{h}_{ij}(x,\rho)$ is defined as in Equ.~(\ref{eq:newh}).

\subsubsection{The Dressed Observables and the Mechanism}
As we have discussed in Sec.~\ref{sec:holography}, physical operators have to satisfy the Hamiltonian and momentum constraints, for which the momentum constraints are easily satisfied as it is imposing the requirement of spatial diffeomorphism invariance. Thus, following Sec.~\ref{sec:holography}, we will study the Hamiltonian constraint in the Karch-Randall braneworld to the first nontrivial order in $G_{N}$. The Hamiltonian constraint is in fact the $00$-th component of the bulk Einstein's equation with matter source. As a result, in our case, the Hamiltonian constraint is the $00$-th component Equ.~(\ref{eq:branegravitoneomsimplify}) with matter source, which is given by
\begin{equation}
\begin{split}
    0=&\sqrt{16\pi G_{N}}e^{4A(\rho)}\bar{g}_{00}\mathcal{H}_{\text{matter}}-\bar{g}_{00}\bar{\nabla}_{m}\bar{\nabla}_{n}\tilde{h}^{mn}+\bar{g}_{00}\text{\ovA{$\Box$}} \tilde{h}-\bar{\nabla}_{0}\bar{\nabla}_{0}\tilde{h}+2\bar{\nabla}_{k}\bar{\nabla}_{0}\tilde{h}_{0}^{k}- \text{\ovA{$\Box$}}\tilde{h}_{00}-(d-1)(\bar{g}_{00}\tilde{h}-2\tilde{h}_{00})\,\\-&e^{-(d-4)A(\rho)}\partial_{\rho}\Big[e^{dA(\rho)}\partial_{\rho}e^{-2A(\rho)}(\tilde{h}_{00}-\bar{g}_{00}\tilde{h})\Big]-e^{-(d-4)A(\rho)}\partial_{\rho}\Big[e^{dA(\rho)}\partial_{\rho}e^{-2A(\rho)}(\bar{\nabla}_{0}V_{0}+\bar{\nabla}_{0}V_{0}-2\bar{g}_{00}\bar{\nabla}_{i}V^{i})\Big]\,,\label{eq:branegravitoneomsimplifysourced}
    \end{split}
\end{equation}
where $\mathcal{H}_{\text{matter}}$ is the bulk matter Hamiltonian density, we have used the fact that $\bar{g}_{0j}=0$ for AdS$_{d}$ and we remember that we stripped off the common factor $e^{-2A(\rho)}$ from the bulk linearized Einstein's equation to get Equ.~(\ref{eq:branegravitoneom}) which was simplified to Equ.~(\ref{eq:branegravitoneomsimplify}). This constraint equation Equ.~(\ref{eq:branegravitoneomsimplifysourced}) can be put into a nice form if we decompose the AdS$_{d}$ metric as
\begin{equation}
    ds^{2}=\bar{g}_{ij}dx^{i}dx^{j}=-N^{2}(x)dt^{2}+\hat{g}_{ab}dx^{a}dx^{b}\,,
\end{equation}
for example in Poincar\'{e} patch $N=\frac{1}{z}$, in terms of which Equ.~(\ref{eq:branegravitoneomsimplifysourced}) becomes
\begin{equation}
    \begin{split}
         0=&\sqrt{16\pi G_{N}}e^{4A(\rho)}\mathcal{H}_{\text{matter}}-\bar{\nabla}_{m}\bar{\nabla}_{n}\tilde{h}^{mn}+\text{\ovA{$\Box$}} \tilde{h}-\bar{g}^{00}\bar{\nabla}_{0}\bar{\nabla}_{0}\tilde{h}+2\bar{g}^{00}\bar{\nabla}_{k}\bar{\nabla}_{0}\tilde{h}_{0}^{k}-\bar{g}^{00} \text{\ovA{$\Box$}}\tilde{h}_{00}-(d-1)(\tilde{h}-2\bar{g}^{00}\tilde{h}_{00})\,\\-&e^{-(d-4)A(\rho)}\partial_{\rho}\Big[e^{dA(\rho)}\partial_{\rho}e^{-2A(\rho)}(\bar{g}^{00}\tilde{h}_{00}-\tilde{h})\Big]+e^{-(d-4)A(\rho)}\partial_{\rho}\Big[e^{dA(\rho)}\partial_{\rho}e^{-2A(\rho)}(-2\bar{g}^{00}\bar{\nabla}_{0}V_{0}+2\bar{\nabla}_{i}V^{i})\Big]\,,\label{eq:branegravitoneomsimplifysourcedfinal}
    \end{split}
\end{equation}
which can be further simplified, as we discussed in Sec.~\ref{sec:holography} by taking $\tilde{h}_{00}$ and $\tilde{h}_{0j}$ to be zero, as
\begin{equation}
    -e^{-4A(\rho)}\Big[(d-2)\hat{h}+\hat{\nabla}_{a}\hat{\nabla}_{b}\hat{h}^{ab}-\hat{\nabla}^{2}\hat{h}\Big]+\sqrt{16\pi G_{N}}\mathcal{H}_{\text{matter}}+e^{-dA(\rho)}\partial_{\rho}\Big[e^{dA(\rho)}\partial_{\rho}e^{-2A(\rho)}(\hat{h}+2\hat{\nabla}^{i}V_{i})\Big]=0\,,\label{eq:Hbrane}
\end{equation}
where $\hat{\nabla}_{a}$ is the metric compatible covariant derivative with respect to the metric $\hat{g}_{ab}$, all upper indices are uplifted from the lower indices using $\hat{g}^{ab}$ and $\hat{h}=\hat{g}^{ab}h_{ab}$. We note that the first term in Equ.~(\ref{eq:Hbrane}) is in the same form as the first term of Equ.~(\ref{eq:core}) which as we have shown in Equ.~(\ref{eq:HADM}) can be written as a total derivative term, and so it will give the ADM Hamiltonian from the brane perspective. Moreover, the third term of Equ.~(\ref{eq:Hbrane}) is in fact the conjugate momentum of the Wilson line $V_{0}(x,\rho)$,
\begin{equation}
     \pi_{V^{0}}(\vec{x},t,\rho)=-e^{-dA(\rho)}\partial_{\rho}\Big[e^{dA(\rho)}\partial_{\rho}e^{-2A(\rho)}(\hat{h}+2\hat{\nabla}^{i}V_{i})\Big]\,,
\end{equation}
which by canonical quantization gives us the equal-time commutator
\begin{equation}
  [\hat{V}^{0}(\vec{y},t,\rho),\hat{\pi}_{V^{0}}(\vec{x},t,\rho')]=i\frac{1}{e^{dA(\rho)}\sqrt{-\bar{g}}}\delta(\rho-\rho')\delta^{d-1}(\vec{y}-\vec{x})\,,
\end{equation}
where we are using $\hat{\quad}$ to denote operators and it is easy to not confuse it with the $\hat{\quad}$ for quantities defined with respect to $\hat{g}_{ab}$.

Now we are ready to study the operators obeying the constraint Equ.~(\ref{eq:Hbrane}). Since we only study local operators, we can equivalently use the integrated version of the constraint Equ.~(\ref{eq:Hbrane}). A scalar operator $\hat{O}(x)$ in the AdS$_{d+1}$ bulk couples to the bulk graviton $h_{\mu\nu}(x,\rho)$, i.e. all the KK gravitons from the brane perspective. It has to satisfy the bulk integrated Hamiltonian constraint
\begin{equation}
    [-\hat{H}_{\partial}^{B}+\sqrt{16\pi G_{N}}\hat{H}_{\text{matter}}-\hat{H}_{B},\hat{O}^{\text{Phys}}(x)]=0\,,
\end{equation}
where we have defined
\begin{equation}
    \begin{split}
        \hat{H}_{\partial}^{B}&=\int e^{dA(\rho)}\sqrt{-\bar{g}} d\rho d^{d-1}\vec{x}e^{-4A(\rho)}\Big[(d-2)\hat{h}+\hat{\nabla}_{a}\hat{\nabla}_{b}\hat{h}^{ab}-\hat{\nabla}^{2}\hat{h}\Big](x,\rho)\\&=\int_{\rho_{B}}^{\infty} e^{(d-4)A(\rho)} d\rho \hat{H}_{\partial}(\rho)\,,\\\hat{H}_{B}&=\int e^{dA(\rho)}\sqrt{-\bar{g}} d\rho d^{d-1}\vec{x}\pi_{V^{0}}(x,\rho)=\lim_{\rho_{c}\rightarrow\infty}\int \sqrt{-\bar{g}}d^{d-1}\vec{x}e^{dA(\rho_{c})}\partial_{\rho}e^{-2A(\rho_{c})}(\hat{h}+2\hat{\nabla}_{i}\hat{V}^{i})(x,\rho_{c})\,,\label{eq:redef}
    \end{split}
\end{equation}
and hence $\hat{H}_{\partial}^{B}$ is a term supported on the defect (i.e. the red dot in Fig.\ref{pic:branedemo}), $\hat{H}_{\text{ADM}}(\rho)=\frac{1}{\sqrt{16\pi G_{N}}}\hat{H}_{\partial}$ is the ADM Hamiltonian for each constant $\rho$-slice and $\hat{H}_{B}$ is supported on the whole leftover asymptotic boundary of the bulk. To obtain the second equation in Equ.~(\ref{eq:redef}), we have used the boundary conditions Equ.~(\ref{eq:bccovariant}). From the point of view of the BCFT dual of the island model in Sec.~\ref{sec:islandmodel}, one can identify $\frac{1}{\sqrt{16\pi G_{N}}}\hat{H}_{\partial}^{B}$ as $\hat{H}_{\text{CFT}_{d-1}}+\hat{H}_{\text{coupling}}$ and $\frac{1}{\sqrt{16\pi G_{N}}}\hat{H}_{B}$ as $\hat{H}_{\text{bath}}$ up to some potential ambiguities near the boundary, where we notice that the island model in this section is $d$-dimensional and that in Sec.~\ref{sec:islandmodel} is $(d+1)$-dimensional. An explicit solution of the constraint, which in fact also solves the momentum constraint, can be constructed as
\begin{equation}
    \hat{O}^{\text{Phys}}(x)=\hat{O}(x+\sqrt{16\pi G_{N}}\hat{V}(x))\,,
\end{equation}
i.e. the arguments are the invariant combinations $x^{\mu}+\sqrt{16\pi G_{N}}\hat{V}^{\mu}(x)$. This is similarly to the situation as demonstrated in Fig.\ref{pic:bulkthetadress}.

Furthermore, from the dual island model perspective, a probe operator in the gravitational AdS$_{d}$ can be thought as coupled only to a few KK gravitons. Let's consider the simplest case that an operator $\hat{\phi}(x)$ only couples to the $n$-th KK graviton. Since this operator sources the $n$-th KK graviton, the operator has to obey the constraint Equ.~(\ref{eq:Hbrane}) projected to the $n$-th KK mode. The integrated version of this constraint is 
\begin{equation}
    [-\hat{H}_{\partial}^{(n)}+\sqrt{16\pi G_{N}}\hat{H}_{\text{matter}}^{(n)}-m_{n}^{2}\int \sqrt{-\bar{g}}d^{d-1}\vec{x}(\hat{h}^{(n)}+2\hat{\nabla}_{i}V^{(n)i}),\hat{\phi}^{\text{Phys}}(x)]=0\,,\label{eq:constraintbrane}
\end{equation}
where we have
\begin{equation}
    \hat{H}_{\text{matter}}^{(n)}\equiv\int \sqrt{-\bar{g}}d^{d-1}\vec{x}d\rho e^{(d-2)A(\rho)}\phi_{n}(x)\mathcal{H}_{\text{matter}}\,,
\end{equation}
which is the Hamiltonian that evolves the probe field $\phi(x)$, and
\begin{equation}
\begin{split}
    \hat{H}^{(n)}_{\partial}\equiv&\int \sqrt{-\bar{g}}  d^{d-1}\vec{x}e^{(d-6)A(\rho)}d\rho\phi_{n}(\rho)\Big[(d-2)\hat{h}+\hat{\nabla}_{a}\hat{\nabla}_{b}\hat{h}^{ab}-\hat{\nabla}^{2}\hat{h}\Big](x,\rho)\,,\\=&\int \sqrt{-\bar{g}}  d^{d-1}\vec{x}\Big[(d-2)\hat{h}^{(n)}+\hat{\nabla}_{a}\hat{\nabla}_{b}\hat{h}^{(n)ab}-\hat{\nabla}^{2}\hat{h}^{(n)}\Big](x)\,,
    \end{split}
\end{equation}
which following the calculation in Equ.~(\ref{eq:HADM}) is exactly the ADM Hamiltonian for the $n$-th KK graviton $\tilde{h}^{(n)}_{ij}(x)$. Moreover, we also notice that the conjugate momentum for the $0$-th component of the $n$-th St\"{u}ckelberg vector field $V^{(n)i}(x)$ is
\begin{equation}
    \pi_{V^{0}}^{(n)}(\vec{x},t)=m_{n}^{2}(\hat{h}^{(n)}+2\hat{\nabla}^{i}V^{(n)}_{i})(\vec{x},t)\,.
\end{equation}
Therefore, we can write Equ.~(\ref{eq:constraintbrane}) as
\begin{equation}
     [-\hat{H}_{\partial}^{(n)}+\sqrt{16\pi G_{N}}\hat{H}_{\text{matter}}^{(n)}-\Pi^{(n)}_{V^{0}},\hat{\phi}^{\text{Phys}}(x)]=0\,.\label{eq:braneconstraintsimpl}
\end{equation}
A solution of Equ.~(\ref{eq:braneconstraintsimpl}) that commutes with the ADM Hamiltonian $\hat{H}^{(n)}_{\text{ADM}}=\frac{1}{\sqrt{16\pi G_{N}}}\hat{H}^{(n)}_{\partial}$ can be constructed as
\begin{equation}
    \hat{\phi}^{\text{Phys}}(x)=\hat{\phi}(x+\sqrt{16\pi G_{N}}\hat{V}^{(n)}(x))\,,\label{eq:KRislando}
\end{equation}
which is a legitimate operator in the entanglement island. It deserves to be mentioned that in the dual island model operators outside the entanglement island should also satisfy Equ.~(\ref{eq:braneconstraintsimpl}) and these operators has to commute with the bath Hamiltonian $\hat{H}_{\partial}$ in Equ.~(\ref{eq:redef}). Such out of island operators can be easily constructed using the gravitational Wilson lines in the AdS$_{d}$ associated with $\tilde{h}^{(n)}_{ij}$ constructed in the same way as Equ.~(\ref{eq:Vdef}) and these operators are evolved by the ADM Hamiltonian $\hat{H}^{(n)}_{\text{ADM}}$ as the result of the constraint. As a check of the consistency condition we proposed in Sec.~\ref{sec:mechanism}, we have for the island operators Equ.~(\ref{eq:KRislando})
\begin{equation}
[\hat{H}_{\text{bath}},\hat{\phi}^{\text{Phys}}(x)]=[\frac{1}{\sqrt{16\pi G_{N}}}\hat{H}_{B},\phi(x+\sqrt{16\pi G_{N}}\hat{V}^{(n)}(x))]=-i\partial_{t}\hat{\phi}^{\text{Phys}}(x)\int_{\rho_{B}}^{\infty}d\rho e^{(d-6)A(\rho)} \phi_{n}(\rho)\neq0\,,
\end{equation}
where we ignored term of higher orders in $G_{N}$ and the $\rho$ integral is finite as normalizable modes decay as $e^{-(d-2)A(\rho)}$ when $\rho\rightarrow\infty$ \cite{Lashkari:2013koa}.\footnote{We only consider $d\geq3$ for which we have graviton in the bulk.} These island operators are similar to those as depicted in Fig.\ref{pic:branethetadress}, i.e. one can think of them as dressed to the bath through the extra dimension to the dual island model in the Karch-Randall braneworld.

\subsection{A Consistency Check with Entanglement Wedge Reconstruction} 
The gravitational Wilson lines that we constructed in Sec.~\ref{sec:gravKR} wrap along the bulk $\rho$-direction i.e. they connect the point on the brane to another point on the asymptotic boundary such that these two points are mirror symmetric with respect to the defect. More explicitly, let's take the empty AdS$_{d+1}$ as the bulk geometry and consider the following coordinate system for the bulk
\begin{equation}
ds^{2}=\frac{1}{\sin^{2}\mu}\Big[d\mu^{2}+\frac{du^{2}+\eta_{ab}dx^{a}dx^{b}}{u^{2}}\Big]\,,
\end{equation}
where $\mu\in(0,\pi)$, $\eta_{ab}$ is the $(d-1)$-dimensional Minkowski metric and the former coordinate $\rho$ is related to $\mu$ as $d\rho=\frac{d\mu}{\sin\mu}$. The branes are constant-$\mu$ slices in this coordinate with the corresponding $\mu_{B}$ determined by the brane tension \cite{Geng:2023qwm}. In this coordinate, the gravitational Wilson lines are wrapping the bulk $\mu$-direction. Therefore, they connect a point $(u,x^{a})$ on the brane to a point on the asymptotic boundary $\mu=\pi$ with the same coordinate $(u,x^{a})$.

A basic consistency condition for the gravitational Wilson lines is that they should stay within the entanglement wedge of the radiation region on the asymptotic boundary if they start inside the island region on the brane. Otherwise the diffeomorphism invariant operator on the brane defined using these gravitational Wilson lines wouldn't be consistent with the entanglement wedge reconstruction as operators in the radiation region. This condition is satisfied if the Ryu-Takayanagi surface starts from the brane point $(u_{B},x^{a})$ ends on the asymptotic boundary $\mu=\pi$ at point $(u_{b},x^{a})$ with $u_{b}\leq u_{B}$. Such a Ryu-Takayanagi surface can be parametrized as $u=u(\mu)$ with $u'(\mu_{B})=0$ \cite{Geng:2020fxl}. This surface is a bulk minimal area surface which obeys the equation
\begin{equation}
u'' = -(d-2) u + (d-1) u' \cot\mu \left(1 - \frac{\tan\mu}{2} \frac{u'}{u} + \frac{u'^2}{u^2}\right)\\
- \left(\frac{d-5}{2}\right) \frac{u'^2}{u}.
\end{equation}
This differential equation satisfies a rather nice property that when $u'=0$ we have
\begin{equation}
    u''=-(d-2)u\leq0\,,
\end{equation}
as we only consider $d\geq2$. Together with the fact that the Ryu-Takayanagi surface satisfies $u'(\mu_{B})=0$, it is not hard to see that $u'(\mu)\leq0$ in the bulk as it goes from the brane to the asymptotic boundary. As a result, we have $u_{b}\leq u_{B}$. Therefore, the above consistency condition is satisfied. We also notice that the above analytic proof can be straightforwardly extended to the case where the bulk geometry is an AdS$_{d+1}$ black string \cite{Geng:2020fxl,Geng:2021mic,Geng:2023qwm}.

\subsection{Potential Implications to the ER=EPR Conjecture}
The ER=EPR conjecture \cite{Maldacena:2013xja} was motivated by the observation that for the eternal black hole in AdS$_{d+1}$ the boundary CFT dual consists of two CFT$_{d}$'s in the thermal-field-double (TFD) state 
\begin{equation}
    \ket{\text{TFD}}=\frac{1}{\sqrt{Z(\beta)}}\sum_{n}e^{-\frac{\beta E_{n}}{2}}\ket{E_{n}}_{L}\ket{E_{n}}_{R}\,,\quad\text{where } Z(\beta)=\sum_{n}e^{-\beta E_{n}}\,,\label{eq:TFD}
\end{equation}
where $\beta<\beta_{HP}$ with $\beta_{HP}$ the inverse Hawking-Page temperature and $\ket{E_{n}}$ are the energy eigenstates of the CFT$_{d}$ \cite{Maldacena:2001kr}. The geometry of the eternal black hole has two asymptotic boundaries and they are geometrically connected in the bulk by an Einstein-Rosen bridge that goes through the bifurcation point. Meanwhile, the dual CFT state in Equ.~(\ref{eq:TFD}) is a highly entangled state between the two CFT's and the AdS/CFT correspondence states that the two CFT's are living on the two asymptotic boundaries of the bulk eternal black hole. Thus, it was conjectured in \cite{VanRaamsdonk:2010pw,Maldacena:2013xja} that the bulk geometry connecting the two CFT's emerges from the entanglement of the state in the dual conformal field theory. This conjecture is supported by many examples of CFT state with known bulk geometric dual. Further studies and considerations such as \cite{Engelhardt:2022qts} suggest that entanglement is not enough to have a connected bulk geometry, even if the entanglement is strong enough. Though, the basic statement that the bulk geometry emerges from features of CFT state is robust and counter-examples to ER=EPR such as \cite{Engelhardt:2022qts} only suggest that the bulk geometry may depend on the features of the CFT state more than those that can be captured by entanglement.

In fact, the mechanism we found for the information encoding scheme of islands and its holographic dual in the Karch-Randall braneworld suggests that it is an explicit example of the ER=EPR conjecture. This can be seen from the following observation. In the Karch-Randall braneworld, the Goldstone vector field in the dual island model is the line integral of the extra-dimensional graviton fields, i.e. the quantum fluctuations of the extra-dimensional metric. Since the Goldstone vector field is the quantum fluctuation of the order parameter for the spontaneously broken diffeomorphism symmetry, one expects that the background extra-dimensional geometry, i.e. the background bulk metric, corresponds to the vacuum expectation value of this order parameter. Interestingly, from the island model point of view, the vacuum expectation value of the order parameter is a feature of the quantum state and as we have seen that it is zero if the AdS and the bath are not coupled. If the AdS and the bath are not coupled, the diffeomorphism on the AdS is not spontaneously broken and so one wouldn't have the Goldstone vector field. Thus the holographic dual for the decoupled case cannot be a higher-dimensional geometry connecting the AdS and the bath like the Karch-Randall braneworld. Otherwise, one would have the line integral of the extra-dimensional graviton fields and its holographic dual would be an extra vector field in AdS that couples to the AdS graviton in the St\"{u}ckelberg form suggesting the spontaneous breaking of the diffeomorphism symmetry in AdS. As we have seen, this line integral of the extra-dimensional graviton field is essentially the gravitational Wilson line. Therefore, from the above analysis, the gravitational Wilson line can be thought of as a microscopic wormhole operator from the point of view of the island model. In other words, the Goldstone vector field in the island model can be thought of as a microscopic wormhole operator which indicates the emergence of a quantum extra dimension that connects the AdS and the bath. When the matter fields in the island model are holographic, this extra dimension becomes classical and the micorscopic wormhole operator becomes the gravitational Wilson line on this classical background.

Last but not least, the fact that the AdS and bath are coupled is important for the above interpretation. One might think that ER=EPR only states that the extra-dimensional geometry emerges from the entanglement between the AdS and the bath so if one decouples the bath and the AdS one can equally have the state with the same entanglement structure as if the AdS and the bath are coupled so the extra dimension should still emerge.\footnote{We thank Daniel Jafferis and Juan Maldacena for bringing this question up.} However, the ground state on the AdS coupled to bath system is in fact not in the Hilbert space of the decoupled system. This is due to the fact that in the coupled case there is strong entanglement between the AdS and the bath near the place they are coupled and this entanglement structure doesn't exist for any state in the Hilbert space of the decoupled system. This entanglement indicates that the AdS and the bath are coupled and so it is essential for the underlying order parameter to have a nonzero expectation value indicating the spontaneous breaking of the AdS diffeomorphism symmetry.

\section{Conclusions and Discussions}\label{sec:conclusions}
In this paper, we uncovered the mechanism which underlies the information encoding of islands by the disconnected non-gravitational system away from it. The essential lesson is that operators in the island are dressed to the non-gravitational bath and one can think of the dressing as to be achieved by line operators going through a highly quantum extra dimension that connects the island and the bath. This picture manifests in the Karch-Randall braneworld for which the matter fields in the dual island model is holographic and so the extra dimensional geometry becomes classical. Moreover, the previous observation that island can exist only in massive gravity \cite{Geng:2020qvw,Geng:2021hlu,Geng:2023ynk,Geng:2023zhq,Geng:2024xpj} plays an important role in this mechanism. The reason is that the spontaneously broken diffeomorphism symmetry in the gravitational spacetime, where the island resides, enables us to have operators in the island that are consistent with the holographic interpretation of the island. That is that operators in the island should commute with operators living in its complement. This is made possible due to the Goldstone vector field that is generated by the spontaneous braking of the diffeomorphism symmetry. Operators in the island can be dressed to obey the diffeomorphism constraints using this Goldstone vector field and the dressed operator is localized inside the island, from the point of view inside the gravitational spacetime. Nevertheless, these dressed operators can be detected by the non-gravitational bath which is the essential mechanism behind the information encoding of island by the non-gravitational bath that we uncovered. We worked out this mechanism in detail in Sec.~\ref{sec:mechanism} for the weakly coupled island model and in Sec.~\ref{sec:toymodel} for a strongly coupled island model with a holographic dual. This is enabled by a more in-depth analysis of both models compared to the early literature \cite{Porrati:2001gx,Porrati:2002dt,Porrati:2003sa,Duff:2004wh,Aharony:2006hz,Karch:2000ct,Karch:2000gx}.

It has been questioned and doubted how the entanglement island resolves the information paradox, as it seems the statement that the information in the island is encoded in the nongravitational bath indicates the existence of some non-locality \cite{Giddings:2021qas,Martinec:2022lsb,Guo:2021blh}. This calls into the question whether the interpretation of island makes sense for local theories. In this work, we found that the above non-locality is perfectly fine with local theories and it is a purely gravitational effects. In fact, in gravitational theories one doesn't expect all observables to be strictly local even though the theory itself is described by a local Lagrangian, at least at the level of an effective field theory. The reason is essentially the gravitational Gauss' law which states that excitations in a gravitational universe can be detected on the surface of a ball-shaped region surrounding this excitation by measuring the gravitational potential. At the quantum level, we expect the energy of the system to be bounded from below and so operators obeying the diffeomorphism constraints in the standard massless gravity should be dressed to the boundary surface which is the quantum manifestation of the gravitational Gauss' law. This is in fact the essence of holography in quantum gravity \cite{Marolf:2006bk,Marolf:2008mf,Marolf:2008tx,Raju:2019qjq, Laddha:2020kvp, Chowdhury:2020hse, Chowdhury:2021nxw, Raju:2021lwh, Chakraborty:2023los}. Even though the island model is atypical as the gravitational theory and the corresponding Gauss' law are modified which allow local operators from the perspective inside the gravitational universe, the essential fact that holography implies the absence of strictly localized observables as a quantum gravitational effect persists. Thus, entanglement island and its holographic interpretation are perfectly fine within local theories of gravity.

\section*{Acknowledgments}
We are grateful to Liam Fitzpatrick, Tom Hartman, Yikun Jiang, Andreas Karch, Daniel Jafferis, Juan Maldacena, Henry Maxfield, Joseph Minahan, Geoff Penington, Suvrat Raju and Lisa Randall for useful discussions. HG would like to thank the hospitality from the Aspen Center for Physics where the final stage of this work is performed. Research at the Aspen Center for Physics is supported by the National Science Foundation grant PHY-2210452 and a grant from the Simons Foundation (1161654, Troyer). The work of HG is supported by a grant from the Physics Department at Harvard University.

\appendix

\section{A Potential Question and Its Answer}\label{sec:AdSCFT}
One might be confused about the first equality in Equ.~(\ref{eq:nonzerocomm1}) as $V^{\mu}(x,z)$ depends on $z$ and $z$ is not necessarily small. 

We should first notice that Equ.~(\ref{eq:nonzerocomm1}) is in fact a summary of our results from the discussions in Sec.~\ref{sec:moreholography}. From Sec.~\ref{sec:moreholography}, we noted that the transformation of $U^{\mu}(x)$ deduced from the bulk large gauge transforms Equ.~(\ref{eq:largegaugeTr}) is proportional to the transformations of $U^{\mu}(x)$ generated by the bath stress-energy tensor as in Equ.~(\ref{eq:Ubath}). Thus, the first equality in Equ.~(\ref{eq:nonzerocomm1}) is summarizing the above observation and saying that \textit{the bulk large gauge transform can be understood as generated by the bath stress-energy tensor}, and it is not suggesting that $V^{\mu}(x,z)=\epsilon^{d+2}U^{\mu}(x)$ for all $z$.

A more careful derivation of this equality is as follows. The transform Equ.~(\ref{eq:Ubath}) suggests that the transformation of the operator $U^{\mu}(x)$ generated by the bath stress-energy tensor can be understood as turning on a constant source proportional to $\epsilon^{\mu}_{\text{bath}}$ for $U^{\mu}(x)$. With this source turned on, the vacuum expectation of the operator $U^{\mu}(x)$ is shifted to
\begin{equation}
    \langle U^{\mu}(x)\rangle \rightarrow\langle U^{\mu}(x)\rangle+\gamma \int d^{d}x \epsilon^{\nu}_{\text{bath}}\langle U^{\mu}(x)U_{\nu}(y)\rangle\,,
\end{equation}
where $\gamma$ is the proportionality constant to ensure that the result is the same as Equ.~(\ref{eq:Ubath}). Using the fact that $U^{\mu}(x)$ has conformal weight $\Delta_{U}=d+3$ and conformal invariance, it is easy to see that the result of the integral must be of the form $\frac{1}{\epsilon^{d+2}}$ with $\epsilon$ as a UV cutoff scale. This shift of the vacuum expectation value due to the sourcing translates to a shift of the operator $U^{\mu}(x)$ if we are considering the original unsourced state. Thus, using AdS/CFT we can see that the effect of the action of the bath stress-energy tensor on the bulk Goldstone vector field $V^{\mu}(x,z)$ is to turn on a constant source for it in the boundary. With the boundary source turned on, its effect on $V^{\mu}(x,z)$ can be decoded using the boundary-to-bulk propagator. Thus, we have
\begin{equation}
    \delta V^{\mu}(x,z)\propto \epsilon^{\nu}_{\text{bath}}\int  d^{d}y K^{\mu}_{\nu}(x,z;y)\,,
\end{equation}
where the boundary-to-bulk propagator $K^{\mu}_{\nu}(x,z;y)$ has been worked out in Section 4 of \cite{Kabat:2012hp}. For the purpose of computing the above integral, the propagator $K^{\mu}_{\nu}(x,z;y)$ is $z\delta^{\mu}_{\nu}$ times the boundary to bulk propagator of a scalar field with 
\begin{equation}
    \Delta=\frac{d}{2}+\sqrt{\frac{(d-2)^2}{4}+2d}=d+1\,,
\end{equation}
where we used Equ.~(62) from \cite{Kabat:2012hp} and the fact that $m^2_{V}=2d$. Thus, we have
\begin{equation}
\begin{split}
    \delta V^{\mu}(x,z)&\propto \epsilon^{\mu}_{\text{bath}}\int d^{d}y\frac{z^{^{d+2}}}{(z^2+(y-x)^2)^{d+1}}\\&=\epsilon^{\mu}_{\text{bath}}\int d^{d}y'\frac{z^{^{d+2}}}{(z^2+y'^2)^{d+1}}\\&\propto\epsilon^{\mu}_{\text{bath}}\int_{0}^{\infty} u^{d-1}du\frac{1}{(1+u^2)^{d+1}}\\&=\epsilon^{\mu}_{\text{bath}}\frac{\sqrt{\pi}\Gamma[\frac{d}{2}]}{2^{d+1}\Gamma[\frac{d+1}{2}]}\,,
    \end{split}
\end{equation}
where the proportionality constant can be fixed by comparing it with the shift of $U^{\mu}(x)$ as in Equ.~(\ref{eq:Ubath}).

In summary, we have proved that the bath Hamiltonian $H_{\text{bath}}$ will induce a constant shift of the bulk Goldstone vector field as described in Equ.~(\ref{eq:nonzerocomm}).

\section{State-dependence and the Petz Map}

There are two interesting questions regarding the relationship between our work and the previous works in black hole physics. The first question regards the state-dependence of operators in the black hole interior \cite{Papadodimas:2011pf,Papadodimas:2012aq,Papadodimas:2013jku,Papadodimas:2015jra,Papadodimas:2015xma,Papadodimas:2017qit}. The second question regards the entanglement wedge reconstruction using the Petz map \cite{Cotler:2017erl,Chen:2019gbt,Penington:2019kki,Bahiru:2022ukn}. Here we articulate these two questions and provide some preliminary discussions to address these two questions.

\begin{displayquote}
\textbf{Question 1}: \textit{It has been understood since the work from Papadodimas and Raju that the operators in the interior of a black hole is state-dependent in AdS/CFT. Entanglement islands could overlap the black hole interior, so one might expect that operators inside the entanglement islands are also state-dependent. How does the mechanism in this work incorporate this state-dependence?}
\end{displayquote}
For this question, we should first emphasize that it is true that entanglement islands could overlap with the black hole interior, but this is not always the case. In fact, as has been understood in \cite{Almheiri:2019yqk,Geng:2021iyq}, with explicit examples, entanglement islands universally exist in the island model, whether or not there is a black hole in the AdS part. Moreover, even in the black hole context, entanglement islands could extend outside the black hole interior \cite{Almheiri:2019yqk,Geng:2020qvw}. Thus, it is too quick to deduce that operators inside the entanglement islands are necessarily state-dependent because operators in the black hole interior are state-dependent in the standard AdS/CFT. More explicitly, the state-dependence of operators in the black hole interior \cite{Papadodimas:2011pf,Papadodimas:2012aq,Papadodimas:2013jku,Papadodimas:2015jra,Papadodimas:2015xma,Papadodimas:2017qit} manifests itself in the bulk and it is due to the sharp black hole horizon. The existence of the black hole horizon indicates that the interior region lies outside the domain of dependence of the exterior region, so if one quantizes a field in the black hole background one should assign different creation and annihilation operators in the interior and the exterior. The operators in the exterior are easily mapped to CFT operators due to the extrapolate dictionary but interior operators are a bit mysterious. However, the smoothness of the horizon requires that the interior and the exterior modes are properly entangled. Thus, there is a relationship between the exterior and interior operators when they are acting on the quantum states. In the decoupling limit of gravity, these quantum states are the perturbative Fock space states built upon a vacuum state associated with the black hole background geometry. Thus, one can see that how the interior operators act is specified by this vacuum state and different black hole microstates should be understood as different such vacuum states \cite{Raju:2020smc}. This is the source of the state-dependence \footnote{For more discussions we refer the readers to \cite{Raju:2020smc}.} and as we can see it is really due to the black hole horizon as a sharp geometric feature. Therefore, it is not so clear how this simple consideration could apply in the context of entanglement islands. We also note that there are arguments suggesting that the state-dependence of the black hole interior operators can be removed if the CFT is coupled to a bath \cite{Verlinde:2013qya}. Hence, we believe that it is an interesting and important question to understand whether the operators inside the entanglement islands are state-dependent.

We should second notice that whether operators inside entanglement islands are state-dependent in the sense of Papadodimas-Raju is in fact orthogonal to the mechanism we uncovered in this paper. This is because our mechanism is a universal gravitational effect, which is independent of the specific manner in which the operators inside the entanglement island are encoded in the bath. This can be easily seen by remembering that the operators in the Papadodimas-Raju construction should also be properly dressed to obey the gravitational constraints and such dressings are done after the fields are quantized and the state-dependence is largely insensitive to this dressing protocol. The mechanism we uncovered is purely due to this dressing and so one naturally expects that it is orthogonal to the specific manner in which the operators inside the islands are encoded in the bath. In fact, with our mechanism one can easily decode operators inside the island using the protocols in \cite{Chowdhury:2020hse}. This decoding works for both state-dependent and state-independent operators. From this perspective, it is similar to the Petz map, but it is more constructive.

As a final remark, we should note that there is in fact a slightly different version of state-dependence in our mechanism. This is because the Goldstone vector boson is really due to the nontrivial AdS to bath correlators. Such correlators indicate fine structures of the entanglement between the AdS and the bath and are features of the quantum state. From this perspective, one can interpret the Goldstone vector boson as state-dependent, and so is our mechanism. We should note that this state-dependence is not obviously related to the Papadodimas-Raju type state-dependence.

\begin{displayquote}
\textbf{Question 2}: \textit{It has been proposed that entanglement wedge reconstruction can be done using the Petz map from the operator algebra quantum error correcting code. How does the mechanism uncovered in this paper relate to this universal reconstruction protocol?}
\end{displayquote}
The Petz map provides a protocol to explicitly reconstruct operators inside the entanglement wedge using the dual CFT operators. Nevertheless, for the protocol to operate, one needs to know the isometry $V$ between the bulk Hilbert space, which is also called the \textit{code subspace}, and the dual CFT Hilbert space.\footnote{In this discussion, we don't review the Petz map in detail. We refer readers to \cite{Cotler:2017erl,Chen:2019gbt,Bahiru:2022ukn} for detailed explanations and exemplified applications.}  In the standard AdS/CFT context, the isometry $V$ is the HKLL reconstruction \cite{Hamilton:2005ju,Hamilton:2006az,Hamilton:2006fh} and the Petz map can be understood as a natural extension from the causal wedge reconstruction to entanglement wedge reconstruction. With the gravitational effects incorporated, the HKLL map $V$ will receive perturbative gravitational
corrections \cite{Kabat:2012av}. However, this doesn't change the formal result that the Petz map is still a reconstruction protocol of the bulk operator by CFT operators, as the gravitational effects only tell us what the correct isometry map $V$ is. This observation naturally extends to the context of entanglement islands. In this context, our mechanism reflects a property of the isometry map $V$ between the island model and the dual BCFT, and it is an intrinsic property of the operator algebra that is independent of a specific reconstruction protocol. This resonates with our discussions at the end of the part for Question 1.

\section{How General the Mechanism Is}\label{sec:general}
From the study in the main text, we can see that we uncovered the mechanism of information encoding of the island by the bath by studying a specific case in which the gravitational region is empty AdS. In this case, all matter fields, including leaky and reflecting sectors, are in the ground state and so there is no matter stress energy source backreacting on the gravitational geometry. Thus, this is a consistent setup whose background configuration obeys all equations of motion. This is the simplest situation for which we know there are islands. The reason we focus on this case is that we want to isolate the intrinsic physics of islands. In this case, the gravitational constraints for operators describing perturbations around this background only have two basic types of solutions:
\begin{itemize}
    \item[\textbf{a)}] Operators dressed to the boundary of the AdS using the gravitational Wilson line;
    \item[\textbf{b)}] Operators dressed to the bath using the Goldstone vector boson $V^{\mu}(x,z)$,
\end{itemize}
and other solutions are linear combinations of them. As we see from the study in the main text, the consistency of the holographic interpretation of islands enforces the choice \textbf{b)} for operators inside the island. Thus, graviton mass is essential for the information encoding of islands by the bath and this relationship manifests even at the level of gravitational perturbation theory due to the perturbatively nonzero commutator Equ.~(\ref{eq:bathresult}).

However, one might wonder how general the mechanism we uncovered in this paper is. For example, what about more complicated background configurations, like an evaporating black hole formed from collapse, for which there are more than the two types of solutions \textbf{a)} and \textbf{b)} for the gravitational constraints. The commutators between the bath Hamiltonian and these other types of solutions are perturbative zero and so are the commutators between the boundary ADM Hamiltonian and these solutions. Thus, one might naively think that it is okay that the commutators between operators inside the island and the bath Hamiltonian are exponentially nonperturbatively small and so the mechanism we uncovered in this paper is a red herring. One might further think that this is consistent with the fact that bath Hamiltonian could be highly chaotic. This is because for chaotic Hamiltonians at finite temperature the spectral gap between nearby states would be $\Delta E\sim e^{-S}<e^{-S_{\text{BH}}}$, where $S$ is the entropy of the bath and it should be larger than the entropy of the black hole $S_{\text{BH}}$ such that one could see a Page curve. Hence, a small excitation would only excite the bath by $\Delta E$ and so the commutator between the operator creating this excitation and the bath Hamiltonian could be as small as $e^{-S_{\text{BH}}}$. 

The resolution for the above puzzle is to notice that entanglement island is a feature of the quantum state not the bath Hamiltonian \cite{Geng:2025byh}. Thus, the mechanism of the information encoding of island by the bath should also depend on the feature of the states that could have island and be independent from a specific choice of the bath Hamiltonian either. As we noticed that the particular feature of states that could support islands is the strong entanglement between the bath and the gravitational universe and this entanglement is exactly the reason for the nonzero graviton mass and the mechanism we uncovered in this paper. However, this consideration only suggests the universality of the mechanism we uncovered. We defer a more careful study which might be based on a case by case analysis to the future.

Before we wrap up, we notice that one might argue that to see the Page curve for a black hole evaporating into a bath one necessarily needs the bath Hamiltonian to be chaotic. In fact, this argument is not correct. This can be seen for example by considering the bath to be described by free fields with a finite geometric size $L$. To see the Page curve, we have to ensure that the degeneracy of the bath states with energy of the same order of the black hole mass $M_{\text{BH}}$ would be larger than the exponential of the entropy of the black hole. For a free bath, the degeneracy can be estimated as the partition number 
\begin{equation}
    p(M_{\text{BH}}L)\sim e^{\sqrt{M_{\text{BH}}L}}\,.
\end{equation}
Thus, to see a Page curve we need
\begin{equation}
    \sqrt{M_{\text{BH}}L} >\frac{A_{\text{BH}}}{4G_{N}}.
\end{equation}
As a result, the spectral gap is
\begin{equation}
   \Delta E\sim \frac{1}{L}\sim \frac{A^2_{\text{BH}}}{M_{\text{BH}}G^2_{N}}\,,
\end{equation}
for which, as an example, if one considers small black holes in four dimensions one has
\begin{equation}
   \Delta E\sim \frac{1}{L}\sim G_N M^3_{\text{BH}}\,,
\end{equation}
which is not exponentially nonperturbatively small.\footnote{In fact, there might be another constraint on the bath spectral gap
\begin{equation}
    \Delta E\sim \frac{1}{L}< T_{\text{BH}}\,,
\end{equation}
which ensures that a particle of Hawking radiation can excite the bath. This is again not exponentially nonperturbatively small.} Hence, there is no obstruction for a non-chaotic bath to see a Page curve and have islands. Thus, we conclude that there is no tension between a perturbatively nonzero spectral gap of the bath Hamiltonian and the Page curve, i.e. islands.

\appendix

\bibliographystyle{JHEP}

\bibliography{references}

\end{document}